%% file: jfm-TEP.tex
\documentclass{jfm}
\usepackage{epstopdf, epsfig}
\usepackage{graphicx}
\usepackage{dcolumn}
\usepackage{bm}
\usepackage[FIGTOPCAP]{subfigure}
\usepackage{CJK}
\usepackage{xcolor}
\usepackage{hyperref}
\usepackage{braket}
\usepackage{float}
\usepackage{lscape}
\usepackage{hyperref}
\usepackage{amsmath}
\usepackage{arcs}
\usepackage{overpic}

\newcommand{\Nc}{N_{\mathcal{C}}}
\newcommand{\Nte}{N_{T\!E}}
\newcommand{\be}{\begin{equation}}
\newcommand{\ee}{\end{equation}}

\newcommand{\blue}{\textcolor{black}}

\newcommand{\SH}{\textcolor{black}}
\newcommand{\JA}{\textcolor{black}}

\shorttitle{Thermoelectric Precession in Turbulent Magnetoconvection}
\shortauthor{Y. Xu, S. Horn and J. M. Aurnou}

\title{Thermoelectric Precession in Turbulent Magnetoconvection}

\author{Yufan Xu\aff{1}
  \corresp{\email{yufanxu@g.ucla.edu}},
  Susanne Horn\aff{1,2}
 \and Jonathan M. Aurnou\aff{1}}

\affiliation{\aff{1}Department of Earth, Planetary, and Space Sciences,
\blue{University} of California, Los Angeles, CA 90095, USA
\aff{2}Centre for Fluid and Complex Systems, Coventry University, \blue{CV1} 2NL, Coventry, UK}

\begin{document}

\maketitle

\begin{abstract}

We present laboratory measurements of the interaction between thermoelectric currents and turbulent magnetoconvection. In a cylindrical volume of liquid gallium heated from below and cooled from above and subject to a vertical magnetic field, it is found that the large scale circulation (LSC) can undergo a slow axial precession. Our experiments demonstrate that this LSC precession occurs only when electrically conducting boundary conditions are employed, and that the precession direction reverses when the axial magnetic field direction is flipped. A thermoelectric magnetoconvection (TEMC) model is developed that successfully predicts the zeroth-order magnetoprecession dynamics. Our TEMC magnetoprecession model hinges on thermoelectric current loops at the top and bottom boundaries, which create Lorentz forces that generate horizontal torques on the overturning large-scale circulatory flow. The thermoelectric torques in our model act to drive a precessional motion of the LSC. This model yields precession \blue{frequency} predictions that are in good agreement with the experimental observations. We postulate that thermoelectric effects in convective flows, long argued to be relevant in liquid metal heat transfer and mixing processes, may also have applications in planetary interior magnetohydrodynamics. 

\end{abstract}



\section{Introduction}\label{sec:intro}
\input{sections/1_intro}

\section{Background}\label{sec:theo_bg}
\input{sections/2_theorybg}

\section{Experimental Set-up and Methods}\label{sec:Exp_method}
\input{sections/3_methods}

\section{Magnetoconvection with Electrically Insulating Boundaries}
\input{sections/4a_MCResults}

\section{Thermoelectric Magnetoconvection with Conducting Boundaries}
\input{sections/4b_TEMCResults}


\section{Thermoelectric Precession Model} 
\label{sec:thermoelectric}
\input{sections/5_TEModel}

\section{Discussion} \label{sec:diss}
\input{sections/6_Discussion}


\clearpage
\appendix
\section{Nondimensional Equations}
\input{sections/Appendix}
\label{nondimApp}

\section{Weakly rotating Magneto-Coriolis Waves}
\input{sections/Appendix2}
\label{MCwave}

\section{Comparison of $\Gamma = 1$ and $\Gamma = 2$ Geometries}
\input{sections/Appendix3}

\label{Gamma1MP}

\section{Data Tables}
\input{sections/data}

\bibliographystyle{jfm}
\bibliography{jfm-bib}


\end{document}

%% file: sections/1_intro.tex
The classical set-up for magnetoconvection (MC) is that of Rayleigh-B\'enard convection (RBC) in an electrically-conductive fluid layer occurring in the presence of an externally imposed magnetic field \citep[e.g.,][]{chandrasekhar1961hydrodynamic, nakagawa1955experiment}. The electrically conducting fluid layer is heated from below and cooled from above, typically with the assumption that the top and bottom horizontal boundaries are isothermal and electrically insulating. The imposed magnetic field is usually vertically- \citep[e.g.,][]{cioni2000, aurnou2001experiments,zurner2020flow} or horizontally-oriented \citep{tasaka2016regular, vogt2018transition}. MC is employed as an idealized model for many physical systems \citep[e.g.,][]{weiss2014magnetoconvection}. In geophysics, MC is considered an essential sub-system of the thermocompositionally driven turbulent convection that generates the magnetic fields in molten metal planetary cores \citep[e.g.,][]{jones2011planetary, roberts2013genesis, aurnou2017cross, moffatt2019}. In astrophysics, MC is associated with the sunspot umbra structure, where the strong magnetic field suppresses the thermal convection in the outer layer of the Sun and other stars \citep[e.g.,][]{proctor1982magnetoconvection, schussler2006magnetoconvection, rempel2009radiative}. MC is also related to the X-ray flaring activities on magnetars with extremely large magnetic flux densities estimated from $10^9$ to $10^{11}\ \mathrm{T}$ \citep{castro2008flares}. Furthermore, MC has an essential role in numerous industrial and engineering applications such as crystal growth \citep[e.g.,][]{moreau1999fundamentals,rudolph2008travelling}, design of liquid-metal-cooled blankets for nuclear fusion reactors \citep[e.g.,][]{barleon1991mhd,abdou2001exploration,salavy2007overview} as well as induction heating, casting \citep[e.g.,][]{taberlet1985turbulent, davidson1999magnetohydrodynamics}, and liquid metal batteries \citep{kelley2018fluid,cheng2021laboratory}.

In sharp contrast to the ideal theoretical MC system, liquid metals employed in many laboratory and industrial MC systems have different thermoelectric properties from the boundary materials. This is also the case in natural systems where the properties \blue{significantly} differ across a material interface, such as at the Earth's core-mantle boundary \citep[e.g.,][]{lay1998, mao2017, mound2019}. When an interfacial temperature gradient is present, thermoelectric currents are generated that can form current loops across the interface \citep[e.g.,][]{shercliff1979thermoelectric, jaworski2010thermoelectric}. When in the presence of magnetic fields that are not parallel to the currents, Lorentz forces arise that can stir the liquid metal \citep{jaworski2010thermoelectric}. Such phenomena can be explained by the thermoelectric magnetohydrodynamics (TE-MHD) theory first developed by \citet{shercliff1979thermoelectric}, which focussed on forced heat transfer in nuclear fusion blankets. Although other applications of TE-MHD exist in solidification processes and crystal growth \citep[e.g.,][]{boettinger2000solidification, kao2009effects}, we are unaware of any previous applications of TE-MHD where the boundary thermal gradients are set by the convection itself \citep[cf.][]{zhang2009}, as occurs in the experiments presented here.

Our laboratory experiment focuses on the canonical configuration of turbulent MC in a cylindrical volume of liquid gallium in the presence of vertical magnetic fields and with different electrical boundary conditions. Three behavioral regimes are identified primarily using sidewall temperature measurements: i) a turbulent large-scale circulation `jump rope vortex (JRV)' regime in the weak magnetic field regime \citep{vogt2018jump};  ii) a magnetoprecessional (MP) regime in which the large-scale circulation (LSC) precesses around its vertical axis is found for moderate magnetic field strengths and electrically conducting boundary conditions; iii) a multi-cellular magnetoconvection (MCMC) regime is found in the highest magnetic field strength cases. Although this is the first \blue{systematic} study of the magnetoprecessional mode, this is not the first time that it has been experimentally observed. This behavior was first observed in our laboratory in the thesis experiments of \citet{grannan2017behaviors}. \blue{In addition, what appears to be a similar precession was reported in the MC experiments of \cite{zurner2020flow}}. 

\blue{The rest of the paper is organized as follows. Section \ref{sec:TEeffects} introduces the fundamentals of thermoelectric effects. Section \ref{sec:GvnEq}} presents the governing equations and non-dimensional parameters that control the TEMC system. Section \ref{sec:nl_sty} reviews the established stability analysis and previous research related to the MC system. Section \ref{sec:Exp_method} addresses the experimental setup, the diagnostics used, and the physical properties of our working fluid, liquid gallium. Section \ref{sec:insulated} shows the experimental results with electrically-insulating boundary conditions. Section \ref{sec:conducting} presents the results of experiments made with electrically-conducting boundary conditions and the appearance of the magnetoprecessional mode. 
Following these laboratory results, in section \ref{sec:thermoelectric}, we develop an analytical model of the magnetoprecessional mode driven by thermoelectric currents generated by horizontal temperature gradients that exist along the top and bottom electrically-conducting boundaries. Finally, Section \ref{sec:diss} contains a discussion of our findings and potential future applications. 

%% file: sections/2_theorybg.tex
\subsection{Thermoelectric Effects}
\label{sec:TEeffects}

Thermoelectric effects enable conversions between thermal and electric energy in electrically conducting materials. There are three different types: the Seebeck, Peltier, and Thomson effects \citep{terasaki20111}. The Peltier and Thomson effects in our experimental system produce temperature changes of order $\mathrm{\mu K}$, which are not resolvable with our present thermometric capabilities. Moreover, such small temperature variations will not affect the dynamics of our system. Thus, Peltier and Thomson effects are not considered further.

The Seebeck effect describes the net spatial diffusion of electrons towards or away from a local temperature anomaly \citep{kasap2001thermoelectric}. As a consequence of this effect, positive and negative charges tend to become sequestered on opposite sides of a regional temperature gradient in the material, leading to the development of a thermoelectric electrical potential. Ohm's law then becomes \citep{shercliff1979thermoelectric}: 
\begin{equation}
\boldsymbol{J} =  \sigma \left( \boldsymbol E + \boldsymbol u \times \boldsymbol B - S \boldsymbol \nabla T \right), 
    \label{equ:seebeck_current}
\end{equation}
where $- \sigma S \boldsymbol \nabla T$ encapsulates the thermoelectric current. The variables in eq. (\ref{equ:seebeck_current}) are the electric current density $\boldsymbol{J}$, the electric conductivity $\sigma$ ($\simeq 3.85 \times 10^6$ S/m in gallium),  the electric field $\boldsymbol E$, the fluid velocity  $\boldsymbol u$,  the magnetic flux $\boldsymbol B$, the Seebeck coefficient $S$, and temperature $T$. 

\citet{mott1958theory} derived the following expression for the Seebeck coefficient of a homogeneous and electrically conducting material as below:
\begin{equation}
    S = -\frac{\pi^{2} {k_B}^{2}x_0}{3 e E_{F 0}} T,
    \label{equ:seebeck}
\end{equation}
where $T$ is measured in Kelvin ($T\approx 300\ \mathrm K$ for room temperature), $k_B = 1.38 \times 10^{-23}\ \mathrm{kg\ m^2 s^{-2} K^{-1}}$ is the Boltzmann constant, $x_0$ is an $O(1)$ dimensionless constant that depends on the material properties, $e = 1.60\times 10^{-19}\ \mathrm C$ is the elementary electron charge, and $E_{F 0}$ is the material's Fermi energy ($\sim 10\ \mathrm{eV} = 1.6 \times 10^{-18}$ J for metals). In a uniform medium, $S$ is a function only of $T$. In this case, $\boldsymbol \nabla S$ is parallel to $\boldsymbol \nabla T$ such that $\boldsymbol \nabla S \times \boldsymbol \nabla T = 0$, which then requires that $S\boldsymbol \nabla T$ is irrotational in a uniform medium.
\begin{figure}
    \centering
  	\includegraphics[width=0.5\textwidth]{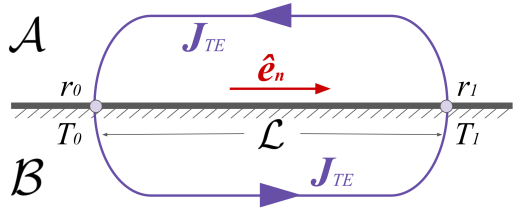}
    \caption{A thermoelectric current loop, $\boldsymbol{J}_{T\!E}$, forms across two different conducting materials $\mathcal{A}$ and $\mathcal{B}$ with a horizontal temperature gradient in thec$\bm{\hat e_n}$-direction. The locations where the thermoelectric current flows in and out the interface are labeled as $r_0$ and $r_1$, respectively. A temperature gradient exists between $r_0$ and $r_1$, where the corresponding temperatures are $T_0 < T_1$. The distance between $r_0$ and $r_1$ is defined as the characteristic length $\mathcal L = |r_1-r_0| $. The direction of the current depends on the Seebeck coefficients of both materials, $S_\mathcal A$ and $S_\mathcal B$, following eq. \blue{(\ref{equ:J_TE})}.}
    \label{fig:TE_AB}%
\end{figure}

As figure \ref{fig:TE_AB} shows, however, a temperature gradient at the interface of two materials with different Seebeck coefficients can generate a net thermoelectric potential. In this case, the Seebeck coefficient $S$ discontinuously varies across the interface of the two materials, $\mathcal A$ and $\mathcal B$. 
Near $r_0$ and $r_1$, $\boldsymbol \nabla S$ is no longer parallel to $\boldsymbol \nabla T$, so a thermoelectric current can form a closed-looped circuit. 

The thermoelectric potential, $\Phi_{T\!E}$, can be calculated via the circuit integral 
\begin{equation}
    {\Phi}_{T\!E}=\oint \frac{\boldsymbol{J}_{T\!E} \cdot d \boldsymbol{r}}{\sigma_{\mathcal{AB}}}, 
    \label{equ:TEemfa}
\end{equation}
where $\sigma_{\mathcal{AB}}$ is the effective electric conductivity of the two-material system. 
The effective electrical resistivity $\tilde{\rho}_{\mathcal{AB}}$ is the sum of the resistivities in each material: 
\begin{equation}
\tilde{\rho}_{\mathcal{AB}} = \tilde{\rho}_\mathcal{\mathcal{A}} + \tilde{\rho}_\mathcal{B} \, ,
\end{equation}
where we assume that the current travels through comparable cross-sectional areas and lengths in each material.
Since $\sigma = 1/\tilde{\rho}$, the effective electrical conductivity for the thermoelectric circuit is 
\begin{equation}
    \sigma_{\mathcal{AB}} =\frac{1}{\tilde{\rho}_{\mathcal{AB}}} 
    = \frac{1}{\Tilde{\rho}_\mathcal{A}+\Tilde{\rho}_\mathcal{B}} = \frac{\sigma_\mathcal{A}\, \sigma_\mathcal{B}}{\sigma_\mathcal{A} + \sigma_\mathcal{B}} \, .
\label{equ:effelec}
\end{equation} 

Isolating the thermoelectric current density in the current loop, $\boldsymbol J_{T\!E}$, in eq. (\ref{equ:seebeck_current}) yields
\begin{equation}
    \boldsymbol{J}_{T\!E} = -\sigma_{\mathcal{AB}} \widetilde S \boldsymbol \nabla T \approx -\sigma_{\mathcal{AB}} \widetilde S \left( \frac{T_1-T_0}{\mathcal L} \right) ,
    \label{equ:J_TE}
\end{equation}
where $\widetilde S$ is the net Seebeck coefficient of the two-material system, and the temperature gradient in the $\bm{\hat e_n}$-direction is approximated by $(T_1-T_0)/{\mathcal L}$ (see figure \ref{fig:TE_AB}). 
Substituting eq. (\ref{equ:J_TE}) into eq. (\ref{equ:TEemfa}), one can show that the net thermoelectric potential ${\Phi}_{T\!E}$ is the difference between the thermoelectric potentials in each material,
\begin{equation}
    {\Phi}_{T\!E}=\Phi_\mathcal{A}-\Phi_\mathcal{B} = -\int_{r_0(T_{0})}^{r_1(T_{1})} S_{\mathcal{A}} \boldsymbol \nabla T \cdot d \boldsymbol r+\int_{r_0(T_{0})}^{r_1(T_{1})} S_{\mathcal{B}} \boldsymbol \nabla T \cdot d \boldsymbol r,
    \label{equ:TEemfb}
\end{equation}
where $r_0$ and $r_1$ denotes the location where the thermoelectric current flows in and out the interface, and $T_0$ and $T_1$ are the temperatures at $r_0$ and $r_1$, respectively. We set $T_0<T_1$, so that the temperature gradient is positive from $r_0$ to $r_1$, following figure \ref{fig:TE_AB}. Here $S_\mathcal{A}$ and $S_\mathcal{B}$ are the Seebeck coefficients of materials $\mathcal{A}$ and $\mathcal{B}$, respectively. 

Substituting eq. (\ref{equ:seebeck}) into eq. (\ref{equ:TEemfb}) then yields
\begin{equation}
    \Phi_{T\!E} = \int_{r_0(T_{0})}^{r_1(T_{1})} (S_{\mathcal{B}}-S_{\mathcal{A}}) \boldsymbol \nabla T \cdot d \boldsymbol r = 
    \frac{\pi^{2} k_B^{2}}{6 e}\left[\frac{x_{\mathcal{B}}}{E_{F \mathcal{B}}}-\frac{x_{\mathcal{A}}}{E_{F \mathcal{A}}}\right] \left(T_1^{2}-T_{0}^{2}\right),
    \label{equ:TEemfc}
\end{equation}
where $x_\mathcal{A}$, $x_\mathcal{B}$, $E_{F\mathcal{A}}$, and $E_{F\mathcal{B}}$ are numerical constants and Fermi energies of the materials $\mathcal{A}$ and $\mathcal{B}$, respectively. 

The system's \textit{net Seebeck coefficient} is then written as 
\begin{equation}
\widetilde{S} = \frac{{\Phi}_{T\!E}}{T_1-T_0} =  \frac{\pi^{2} \blue{k_{B}}^{2}}{3 e}\left[\frac{x_{\mathcal{B}}}{E_{F \mathcal{\mathcal{B}}}}-\frac{x_{\mathcal{A}}}{E_{F \mathcal{A}}}\right] \left( \frac{T_0+T_1}{2} \right),
    \label{equ:SeebeckAB}
\end{equation}
where $(T_1+T_0)/2$ is the mean temperature of the material interface. Note the structural similarity between the expressions for the Seebeck coefficient for a single material eq. (\ref{equ:seebeck}) and the net Seebeck coefficient across a material interface eq. (\ref{equ:SeebeckAB}).

\subsection{Governing Equations and Nondimensional Parameters}
\label{sec:GvnEq}

The magnetic Reynolds number, $Rm$, estimates the ratio of magnetic induction and diffusion in an MHD system. In our laboratory experiments, upper bounding values of $Rm$ are estimated by using the convective free-fall velocity \citep{julien1996rapidly,glazier1999evidence},  
\be 
U_{f\!f} =\sqrt{\alpha_T \Delta T g H } \, ,
\label{Eq:ff}
\ee
leading to 
\begin{equation}
    Rm = \frac{U_{f\!f} H}{\eta} = Re \, Pm \, ,
\end{equation}
\JA{where $Re$ is the Reynolds number, which denotes the ratio of inertial and viscous effects,
\begin{equation}
    Re = \frac{U_{f\!f} H}{\nu} \, ,
\end{equation}
the magnetic Prandtl number is the ratio of the fluid's magnetic diffusivity $\eta$ and its kinematic viscosity $\nu$, 
 \begin{equation}
    Pm = \frac{\nu}{\eta} \, ,
\end{equation}
and} $\alpha_T$ is the thermal expansivity of the fluid, $\Delta T$ is the vertical temperature difference across the fluid layer of depth $H$, $g$ is the gravitational acceleration.
In our experiments, $Re \lesssim 9 \times 10^3$ and $Pm \simeq 1.7 \times 10^{-6}$.  Thus, $Rm \lesssim 0.015 \ll 1$ for our system, in \JA{good} agreement with estimates made using ultrasonic velocity measurements in this same \JA{setup} by \citet{vogt2018jump}.  Further, the free-fall timescale can be defined as
\be 
\tau_{f\!f} = H / U_{f\!f}\, .
\ee

The estimates above show that magnetic diffusion dominates induction in our experiments. In this low-$Rm$ regime, the influence of fluid motions on the magnetic field can be neglected and the \JA{full} magnetic induction equation need not be solved amongst the governing equations. This results in both $Rm$ and $Pm$ dropping out of the problem \citep{davidson_2016}. This so-called `quasistatic approximation' is commonly applied in low-$Rm$ fluid systems and is valid in most laboratory and industrial liquid metal applications \citep[e.g.,][]{sarris2006limits, davidson_2016,knaepen2008magnetohydrodynamic}. 

In addition to quastistaticity, the Boussinesq approximation is applied \citep{oberbeck1879warmeleitung, boussinesq1903theorie, gray1976validity, Tritton77, chilla2012new} and the governing equations of thermoelectric magnetoconvection (TEMC) are then
\begin{subequations}
\label{dimGE}
\begin{equation} 
	 \boldsymbol{J} = \sigma \left( - \boldsymbol \nabla \Phi  + \boldsymbol{u} \times \boldsymbol{B} - S \boldsymbol \nabla T\right), 
	 \label{equ:ohmslaw} 
\end{equation}	
\begin{equation}
	 \boldsymbol \nabla  \boldsymbol \cdot  \boldsymbol { J } = 0,
	 \label{equ:chargeconservation}
\end{equation}	 
\begin{equation}
    \frac { \partial \boldsymbol{u} } { \partial t } + ( \boldsymbol{u} \cdot \boldsymbol  \nabla ) \boldsymbol{u}  =  
    - \frac{1}{\rho } \boldsymbol \nabla  p  + \frac { 1 } { \rho } ( \boldsymbol{J} \times \boldsymbol{B} ) + \nu {\boldsymbol  \nabla}^{ 2 } \boldsymbol{u} + \alpha _ { T } \Delta T \boldsymbol{g},
    \label{equ:momentum_gov}
\end{equation}	 
\begin{equation}
	 \boldsymbol \nabla  \boldsymbol \cdot  \boldsymbol { u } = 0,
	 \label{equ:continuity}
\end{equation}	
\begin{equation}
    \frac { \partial T } { \partial t } + (\boldsymbol{u} \cdot \boldsymbol  \nabla ) T = \kappa {\boldsymbol \nabla}^{ 2 } T,
    \label{equ:energy_gov}
\end{equation}	 
\end{subequations}
where $\rho$ is fluid density, $p$ is non-hydrostatic pressure, \JA{$\boldsymbol g = g \bm{\hat e_z}$} is the gravity vector, and $\kappa$ is the thermal diffusivity. The external field is $\boldsymbol B = B \bm{\hat e_b}$. Note that Ohm's law (\ref{equ:ohmslaw}) has been simplified via the quasistatic approximation, such that the rotational part of electric field and perturbative second-order terms from $\boldsymbol u \times \boldsymbol{B}$ are not considered. Accordingly, in the bulk fluid, far from material interfaces, where net Seebeck effects are small, the quasistatic Lorentz force is $\boldsymbol J \times \boldsymbol B \sim -\sigma \boldsymbol u_{\perp} B^2$, where $\boldsymbol u_{\perp}$ is the velocity perpendicular to the direction of the magnetic field. Therefore, the low-$Rm$ Lorentz force acts as a drag that opposes bulk fluid velocities that are directed perpendicular to $\boldsymbol{B}$ \citep{sarris2006limits, davidson_2016}. This quasistatic Lorentz drag depends only on $B^2$. In sharp contrast, the thermoelectric component of the Lorentz force, $- \sigma S \boldsymbol \nabla T \times \boldsymbol{B} $, varies linearly with $\boldsymbol{B}$. Therefore, the thermoelectric Lorentz force changes sign when the direction of the applied magnetic field is flipped.

The dimensionless form of the TEMC governing equations are given in (\ref{nondimGE}) and (\ref{nondimFL}) in Appendix \ref{AppA}. The nondimensional control groups in Appendix \ref{AppA} may be decomposed into four 
parameters: the Prandtl number $Pr$, the Rayleigh number $Ra$, the Chandrasekhar number $Ch$, and the Seebeck number $Se$. The Prandtl number describes the thermo-mechanical properties of the fluid:
\begin{equation}
    Pr = \frac{\nu}{\kappa};
\end{equation}
in liquid gallium, $Pr \approx 0.027$ at $40^{\circ} \mathrm C $. 
The Rayleigh number characterizes the buoyancy forcing relative to thermoviscous damping: 
\begin{equation}
   Ra = \frac{\alpha_T \Delta T g H^3}{\nu \kappa}.
\end{equation}
The Chandrasekhar number describes the ratio of quasistatic Lorentz and viscous forces:
\begin{equation}
   Ch = \frac{\sigma B^2 H^2}{\rho \nu}.
\end{equation}
The Seebeck number estimates the ratio \JA{of} thermoelectric currents in the fluid and \JA{currents} induced by fluid motions: 
\begin{equation}
    Se =  \frac{\blue{|\widetilde{S}|} \Delta T/H}{U_{f\!f} B }  \, . 
    \label{equ:defSe}
\end{equation}
Alternatively, $Se$ can be cast as the ratio of the thermoelectrical potential and the motionally-induced potential in the fluid. Typical values of $Se$ in our experiments with gallium-copper interfaces range from $O(10^{-2})$ to $O(1)$, implying that the Seebeck effect can generate dynamically significant experimental thermoelectric currents.

Lastly, the aspect ratio acts to describe the geometry of the fluid volume: 
\begin{equation}
   \Gamma = \frac{D}{H},
\end{equation}
where $D$ is the inner diameter of the cylindrical container. We focus on $\Gamma = 2$ in this study, similar to \citet{vogt2018jump}, \JA{and present only two $\Gamma = 1$ case results for contrast in Appendix \ref{Gamma1MP}.} 
\begin{table}
    \centering
    \def~{\hphantom{0}}
    \begin{tabular}{l c c c c c}
\textbf{Number Names} & \textbf{Symbol} & \textbf{Definition}  & \textbf{Equivalence} & \textbf{Current Study}\\ \hline
Magnetic Reynolds     & $Rm$      & $\displaystyle \frac{U_{f\!f}H}{\eta}$                     & $RePm$    &    $\lesssim 10^{-2}$                 \\[12pt]
Magnetic Prandtl      & $Pm$      & $\displaystyle \frac{\nu}{\eta}$              &    &  $1.7\times 10^{-6}$            \\ \hline
Prandtl               & $Pr$      & $\displaystyle \frac{\nu}{\kappa} $        &    &  $2.7\times 10^{-2}$                         \\[12pt]
Rayleigh              & $Ra$      & $\displaystyle \frac{\alpha g \Delta T H^3}{\nu \kappa}$  &  &    $\sim 2\times 10^6$\\[12pt]
Chandrasekhar         & $Ch$      & $\displaystyle \frac{\sigma B^2 H^2}{\rho \nu}$     &       & $[0, 8.4\times 10^4]$                       \\[12pt]
Seebeck               & $Se$       & $\displaystyle{\frac{\blue{|\widetilde{S}|}\Delta T/H}{U_{f\!f} B}}$   &  & $ \sim [10^{-2}, 1]$ \\[12pt]
Aspect Ratio          & $\displaystyle \Gamma$  & $\displaystyle \frac{D}{H}$                  &   & $2.0$  \\ \hline
Reynolds    & $Re$      & $\displaystyle \frac{U_{f\!f}H}{\nu}$           & $\displaystyle \sqrt{\frac{Ra}{Pr}}$    & $ \lesssim 8.7\times 10^3$ \\[12pt] 
P\'eclet    & $Pe$      & $\displaystyle \frac{U_{f\!f}H}{\kappa}$    & $\displaystyle \sqrt{Ra Pr}$          & $ \lesssim 2.2\times 10^2$               \\[12pt]  
Convective Interaction & $\Nc$       & $\displaystyle \frac{\sigma B^2H}{\rho U_{f\!f}}$   &  $\displaystyle \sqrt{\frac{Ch^2 Pr}{Ra}} = \frac{Ch}{Re}$  & $\lesssim 10$ \\[12pt]   
Thermoelectric Interaction & $\Nte$       & $\displaystyle \frac{\sigma B \blue{|\widetilde{S}|} \Delta T}{\rho U_{f\!f}^2}$   &  $\displaystyle Se N_{\mathcal{C}} $  & $\lesssim 10$ \\[12pt]  
    \end{tabular}
    \caption{Nondimensional parameters and parameter groups in thermoelectric magnetoconvection (TEMC). The low values of the top two parameters show that the current experiments fall within the quasistatic approximation.  The next five are the base parameters used to describe most of the experimental \JA{cases}. The \blue{next} four parameters are alternative groupings that arise in the 
\JA{nondimensional version of eq. (\ref{dimGE}) given in Appendix \ref{AppA}. }
    }
    \label{table:NDparameters}
\end{table}

Alternatively, the groups of the above parameters that exist in (\ref{nondimGE}) and (\ref{nondimFL}) are the P\'eclet number $Pe$, the Reynolds number $Re$, the convective interaction parameter $N_{\mathcal C}$ and the thermoelectric interaction parameter $N_\mathcal{C}Se$. 
The P\'eclet number,
\begin{equation}
Pe = \frac{U_{f\!f} H}{\kappa} = \sqrt{Ra Pr},
\end{equation}
estimates the ratio of thermal advection and thermal diffusion in the thermal energy equation.  The convective interaction parameter $N_{\mathcal C}$ is the ratio of quasistatic Lorentz drag and fluid inertia. It is defined as: 
\begin{equation}
    N_{\mathcal{C}} = \frac{\sigma B^2H}{\rho U_{f\!f}} = Ch\sqrt{\frac{Pr}{Ra}} = \frac{Ch}{Re}.
    \label{equ:defNc}
\end{equation}
When $N_\mathcal{C}\gtrsim 1$, the Lorentz force will tend to strongly damp buoyancy-driven convective turbulence. Lastly, the thermoelectric interaction parameter, $N_{T\!E}$ is the product of the convective interaction parameter $N_{\mathcal C}$ and the Seebeck number $Se$. 
This parameter approximates the ratio between the thermoelectric Lorentz force and the fluid inertia, and is given by
\begin{equation}
    \Nte = \frac{\sigma B \blue{|\widetilde{S}|} \Delta T}{\rho U_{f\!f}^2} = Se \, N_{\mathcal{C}} .
    \label{equ:defNTE}
\end{equation}
Thus, when $Se \sim 1$, the thermoelectric forces can become comparable to the MHD drag, at least in the vicinity of the material interfaces where the thermoelectric currents are maximal. 

All the nondimensional parameters and their estimated values for our study are summarized in Table \ref{table:NDparameters}. 

\subsection{Previous Studies of Turbulent Magnetoconvection}
 \label{sec:nl_sty}
Despite its broad relevance to natural and industrial systems, magnetoconvection has not been studied in great detail relative to non-magnetic RBC \citep[e.g.,][]{Ahlers09} and rotating convection \citep[e.g.,][]{Aurnou15}. Further, laboratory and numerical studies of turbulent MC have largely neglected thermoelectric effects to date \citep[cf.][]{zhang2009}. Thus, in reviewing the current state of turbulent MC studies, TE effects will not be considered. 

In the limit of weak magnetic fields, such that $N_{\mathcal{C}} \rightarrow 0$, turbulent MC behaves similarly to RBC \citep{cioni2000, zurner2016heat}, with the flow self-organizing into a large-scale circulation (LSC). Thus, the LSC is the base flow structure in turbulent MC when the dynamical effects of the magnetic field are subdominant \citep{zurner2020flow}. LSCs, the largest turbulent overturning structure in the bulk fluid, have been studied extensively in RBC systems \blue{\citep[e.g.,][]{xia2003particle, xi2004laminar, sun2005three, von2008large, brown2009origin, Ahlers09, chilla2012new, pandey2018turbulent, stevens2018turbulent, vogt2018jump, zurner2020flow}.} 

\JA{\citet{vogt2018jump} carried out turbulent RBC laboratory (and associated numerical) experiments in a $\Gamma = 2$ liquid gallium cell using the same laboratory device as we employ in this study. Coupling the DNS outputs to laboratory thermo-velocimetric data, \citet{vogt2018jump} found that the \blue{turbulent} liquid metal convection was dominated by a so-called jump rope vortex (JRV) LSC mode, \blue{instead of the sloshing and torsional modes found in the majority of $\Gamma = 1$ experiments \citep[e.g.,][]{funfschilling2004plume, funfschilling2008torsional, brown2009origin,xi2009origin,zhou2009oscillations}}.} The JRV had a characteristic oscillation frequency $\widetilde f_{JRV}$ of
\begin{equation}
    \widetilde f_{JRV} = f_{JRV}/f_\kappa = 0.027Ra^{0.419},
    \label{equ:jrv}
\end{equation}
where $f_\kappa$ is the inverse of the thermal diffusion timescale
\begin{equation}
    \tau_\kappa = H^2 / \kappa.
\end{equation}
\JA{Ultrasonic} measurements yielded an LSC velocity scaling corresponding to 
\begin{equation}
    Re_{JRV} = 0.99\left({Ra}/{Pr} \right)^{0.483},
    \label{equ:ReRa}
\end{equation}
formulated using their mean Prandtl number value, $Pr \simeq 0.027$. These velocity measurements approach the free-fall velocity scaling in which $Re = U_{f\!f} H / \nu = (Ra/Pr)^{1/2}$.  Thus, we will use $U_{f\!f}$ as the characteristic velocity scale when nondimensionalizing our equations in Appendix \ref{nondimApp} and in the model of thermoelectric LSC precession developed in \S \ref{sec:thermoelectric}. 

The quasistatic Lorentz force does, however, impede the convective motions in finite $N_{\mathcal{C}}$ cases.  \citet{zurner2020flow} used ultrasonic velocimetry measurements to develop an empirical scaling law for the global characteristic velocity, $U_{MC}$, in GaInSn MC experiments:
\begin{equation}
    U_{MC} =  \left( \frac{1}{1+0.68 \, \Nc^{0.87}} \right) \, U_{f\!f}.
    \label{equ:Zurner}
\end{equation}
In \S \ref{sec:thermoelectric}, we will test both $U_{MC}$ and $U_{f\!f}$ in our model for thermoelectrical precession of the LSC, and show that the $U_{MC}$-based predictions better fit our precessional frequency measurements. 

The turbulent LSC mode breaks down in MC when $\Nc \gtrsim 1$ \citep{cioni2000, zurner2019combined, zurner2020flow}. This is roughly analogous to the loss of the LSC in rotating convection when the Rossby number is decreased below unity \citep{kunnen2008breakdown, horn2015toroidal}. In the supercritical $\Nc \gtrsim 1$ regime, the convection in the fluid bulk should then become multi-cellular, akin to the flows shown in \citet{yan2019heat}.  

Near the onset of the magnetoconvection, wall modes appear near the vertical boundaries and will become unstable before bulk convection in many geometrically-confined MC systems \citep{busse2008asymptotic}.

It is important to stress that MC wall modes do not drift along the wall, in contrast to rotating convection \citep{ecke1992hopf}, since the quasistatic Lorentz force does not break azimuthal reflection symmetry \citep[e.g.,][]{houchens2002rayleigh}. The multi-cellular and magneto-wall mode regimes were both investigated in the numerical MC simulations of \citet{liu2018wall}.  The wall modes were found not to drift in their large-aspect ratio simulations, similar to the experimental findings of \citet{zurner2020flow}.  Further, \citet{liu2018wall} showed that the wall modes could become unstable and inject nearly axially-invariant jets into the fluid bulk.  

Strong wall mode injections are also found in the numerical MC simulations of \citet{akhmedagaev2020turbulent}.  These injected axially-invariant jets are accompanied by a net azimuthal drift of the flow field, whose drift direction appears to be randomly set. We interpret these drifting flows as being controlled by the collisional interaction of the jets, qualitatively similar in nature to the onset of the shearing flows in the plane layer simulations of \citet{goluskin2014}. Therefore, we argue that the drifting effect found in the $\Nc > 1$ near-onset numerical simulation by \citet{akhmedagaev2020turbulent} fundamentally differs from the LSC precession found in the thermoelectrically-active $\Nc \lesssim 1$ experiments reported herein.

%% file: sections/3_methods.tex
Laboratory MC experiments are conducted using UCLA's RoMag device, as shown in figure \ref{fig:apparatus}. \blue{See the appendix} of \citet{king2012} for device details. Here, a vertical magnetic field is applied to an upright, non-rotating cylindrical tank filled with liquid gallium ($Pr \simeq 0.027$). The magnetic field vector is 
\be
\bm{B} = B \bm{\hat e_b}, \quad \mbox{where} \quad \ \bm{\hat e_b} = \pm \bm{\hat e_z}, 
\ee
such that $\bm{\hat e_b} = + \bm{\hat e_z}$ corresponds to an upward magnetic field vector and $\bm{\hat e_b} = - \bm{\hat e_z}$ corresponds to a downward magnetic field vector. 
The magnetic field is generated by an hourglass solenoid. With the tank centered along the bore of the solenoid, the vertical component of the magnetic field is constant over the fluid volume to within $\pm 0.5$\% \citep{king2015magnetostrophic}. The magnetic field strength can be varied from 0 to 650 gauss, corresponding to a maximum Chandrasekhar value of $Ch = 8.4 \times 10^4$.

The material properties of liquid gallium are adapted from \citet{aurnou2018rotating}. The container is made up of a cylindrical sidewall and a set of top and bottom end-blocks; the sidewall has an inner diameter $D = 2R = 196.8\ \mathrm{mm}$ and the fluid layer height is fixed at $H = 98.4\ \mathrm{mm}$ such that $\Gamma= 2.0$. We control the thermoelectric effects by changing the materials of these bounding elements. In specific, two different sets of boundaries are used. The first set is made up of an acrylic sidewall and Teflon coated aluminum end-blocks, in order to achieve electrically-insulated boundary conditions. The second set uses a stainless steel 316L sidewall and copper end-blocks, which provide electrically-conducting boundary conditions. The copper is uncoated and has been allowed to chemically interact with the gallium. \blue{This copper interface is not perfectly smooth due to gallium corrosion, allowing gallium to fully wets the surface.} This is important as liquid metals often fail to make good surface contact with extremely smooth, pristine surfaces, likely due to strong surface tension effects. 

The bottom of the convection stack is heated with a non-inductively wound electrical resistance  pad (figure \ref{fig:apparatus}(c)), with the heating power held at a fixed value, $P_{input}$, in each experiment. Heat is extracted at the top of the convection stack by circulating thermostated cooling fluid through an aluminum heat exchanger that contains a double-spiral internal channel. Although the double wound channel minimizes the temperature gradients within the heat exchanger, the inlet and outlet ports must be at different temperatures due to the extraction of heat from the tank. The locations of the cooler inlet and warmer outlet are marked by arrows and triangles in figure \ref{fig:apparatus}(d), and in later figures just by the triangles.

By maintaining the time-mean difference between the horizontally averaged temperatures on the top and bottom boundaries, we are able to fix $Ra \approx 2\times 10^6$ for all the experiments in this study. The sidewall of the tank is thermally insulated by a 5-cm thick Aspen Aerogels' Cryogel X201 blanket (not shown), which has a thermal conductivity of  $0.015\ \mathrm{W/(m\, K)}$. The heating power lost from the sidewall and endwalls, $P_{loss}$, is estimated and then subtracted from the total input power, so that the effective heating power is $P = P_{input}-P_{loss}$.  

Twelve thermistors are embedded in the top and bottom end-blocks roughly \blue{$\delta z = 2 \, \mathrm{mm}$} from the fluid-solid interface, and \blue{at cylindrical radius $r = 0.71 R$}. These are shown as the red probes in figure \ref{fig:apparatus}(d)). These thermistors are evenly separated $60^\circ$ apart from each other in azimuth. Another six thermistors, shown in green in figure \ref{fig:apparatus}(d), are located on the exterior wall of the sidewall in the tank's midplane. The midplane thermistors are located at the same azimuth values as the top and bottom block thermistors, forming six vertically aligned thermistor triplets. Temperature data are simultaneously acquired at a rate of $10\, \mathrm{Hz}$. 

\begin{figure}
    \centering
  	\includegraphics[width=\textwidth]{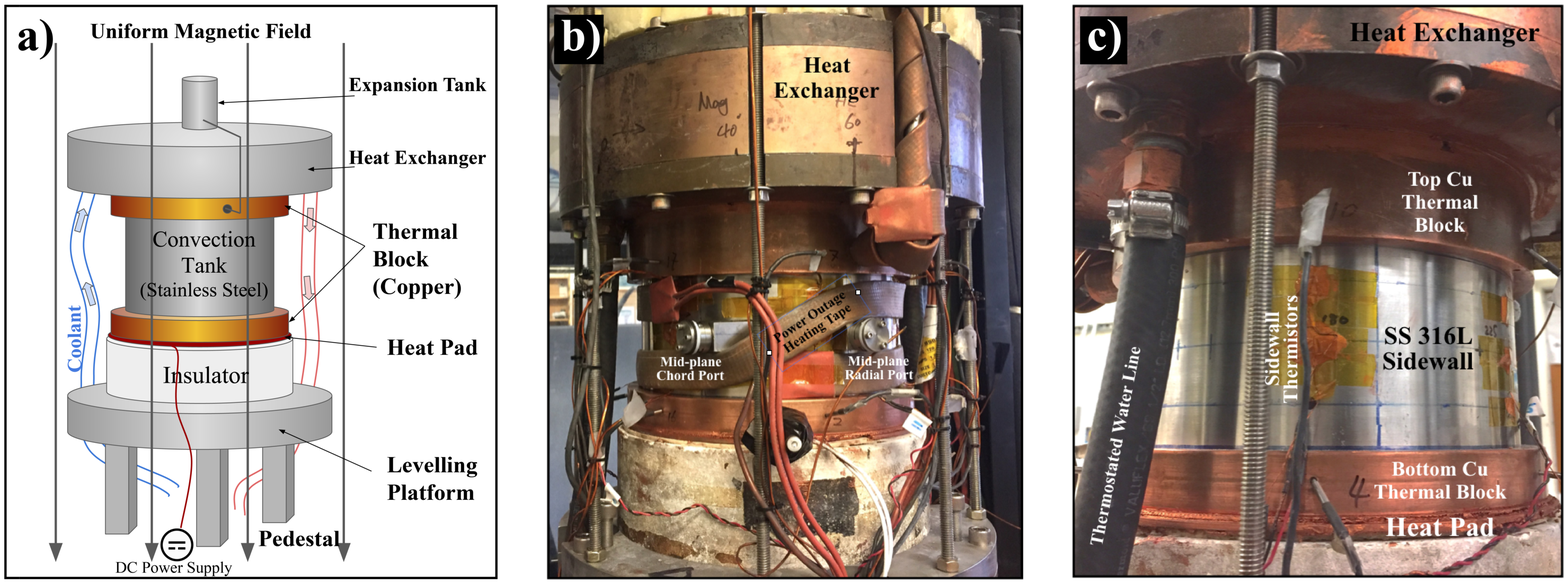}\\ 
  	\vspace{10pt}
  	\includegraphics[width=\textwidth]{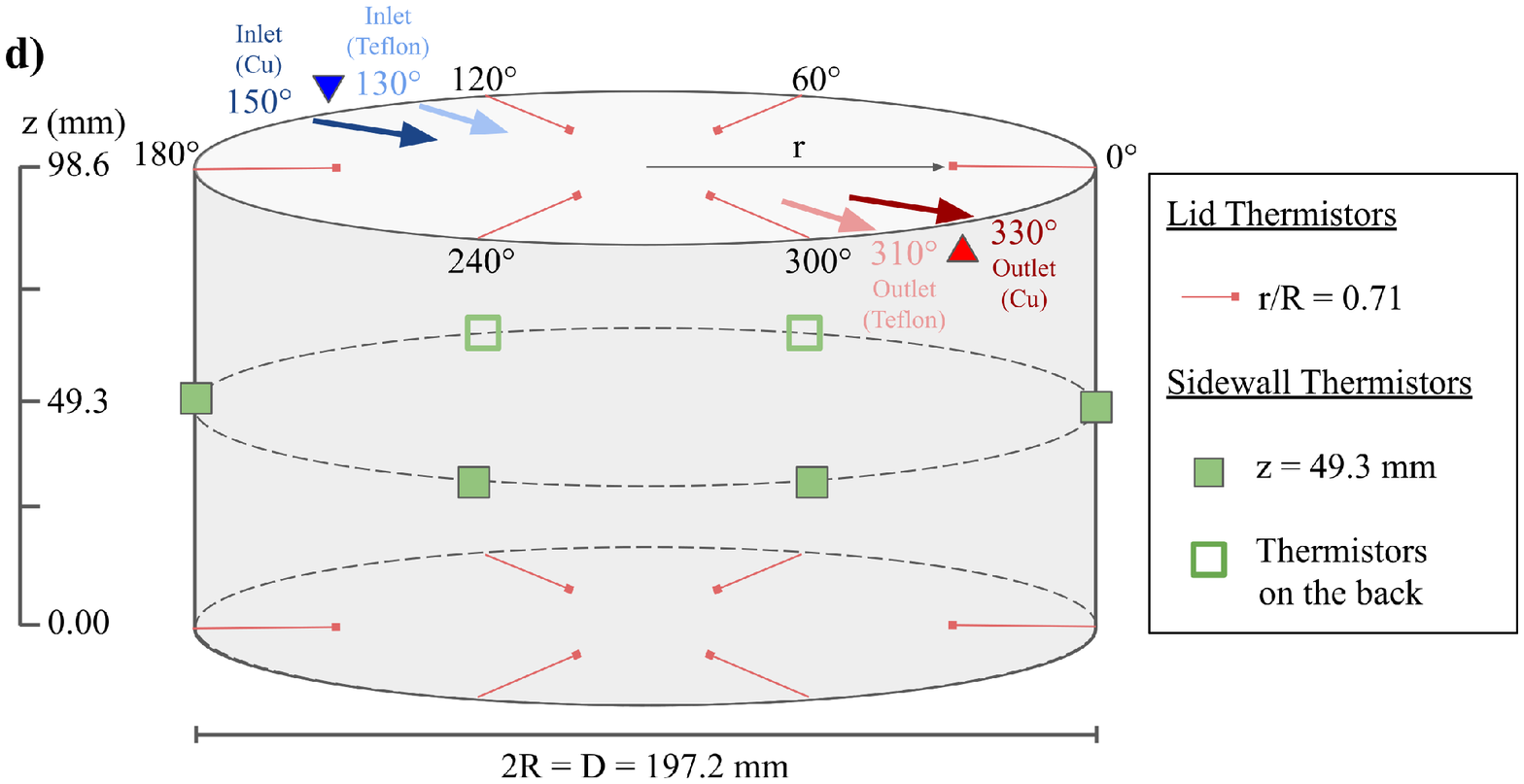}
\caption{(a)  Schematic of the laboratory apparatus. (b)  Image of the convection tank with heat exchanger and safety heating tape in case of \blue{a} power outage. (c) Closer in image of the sidewall and top and bottom thermal end-blocks. The device is thermally insulated by an aerogel blanket that is not shown here. (d) Schematic showing the top, bottom, and midplane thermistor placements. The sidewall midplane thermistors vertically align with the top and bottom thermistor locations. The top and bottom thermistors are located $2\ \mathrm{mm}$ from the fluid surfaces and extend horizontally $28.9\ \mathrm{mm}$ into the lids from the side. The blue and red arrows on \blue{the} top mark the azimuth position of the inlet (cooler coolant) and outlet (warmer coolant) locations on the heat exchanger. In \blue{the} following figures, these azimuthal angles are marked by the downward blue triangle and the upward red triangle.}
\label{fig:apparatus}%
\end{figure}

This thermometry data is discretized in both space and time.  The discrete temperature time series data is expressed as  
\begin{equation}
    T^k_{ij} = T(\phi_i,t_j,z_k).
 \label{equ:Tijk}
\end{equation}
The index $i$ ranges from $1$ to $6$, corresponding to the thermistor locations at $0^{\circ}$, $60^{\circ}$, $120^{\circ}$, $180^{\circ}$, $240^{\circ}$, and $300^{\circ}$ azimuth, respectively. The time step in the data acquisition is denoted by the index $j$, which ranges from 1 to a final index value $N$ for a given time series. Thermistor height is labeled via index $k = 1,\, 2$, or $3$, corresponding to the bottom block thermistors, the midplane thermistors and the top block thermistors, respectively. The bottom block thermistors are located at $z_1 = -2\ \mathrm{mm}$ = `$bot$'; the midplane thermistors are at $z_2 = 49.3\  \mathrm{mm} = $ `$mid$'; and the top block thermistors are set at $z_3 = 100.6\ \mathrm{mm} = $ `$top$'.  No index is given for the radial position of the thermistors, so we reiterate that the end-block thermistors ($k = 1, \, 3$) are located at $r = 69.7 \ \mathrm{mm} = 0.71 R$, whereas the midplane thermistors ($k = 2$) are on the exterior of the sidewall at $r = 100 \ \mathrm{mm} = 1.02 R$.

The thermometry data is used to calculate the time-averaged temperature difference across the height of the fluid layer, $\Delta T$, defined as 
\be
\Delta T =  T^{bot} - T^{top} \, , 
\label{Eq:DeltaT}
\ee
where  $T^{bot}$ and $T^{top}$ are the time and azimuthal mean temperatures of the bottom and top end-block boundaries. These horizontal means are calculated via 
\begin{equation}
T^{bot} = \frac{1}{6N}  \sum_{i=1}^6 \sum_{j=1}^N  T^{bot}_{ij}  \quad \mbox{and} \quad T^{top} = \frac{1}{6N}  \sum_{i=1}^6 \sum_{j=1}^N  T^{top}_{ij}  \, . 
\label{Eq:HorizMean}
\end{equation}
This indexing convention will be used throughout this treatment. Further, $T^k$ denotes the time-azimuthal mean temperature on the index $k$ horizontal plane. 

The material properties of the working fluid are determined using the mean temperature of the fluid volume 
\be
\overline{T} = \left( T^{bot} + T^{top} \right) / 2 \, .
\label{Eq:MeanT}
\ee 
These, in turn, can then be used to measure the heat transfer efficiency of the system, characterized by the Nusselt number, 
\begin{equation}
    Nu = \frac{qH}{\lambda \Delta T},
\end{equation}
where $q = 4 P/(\pi D^2)$ is the heat flux, and $\lambda = 31.4 \ \mathrm{W/(m\, K)}$ is the thermal conductivity of gallium. The Nusselt number describes the ratio of the total and conductive heat transfer across the fluid layer \citep[e.g.,][]{cheng2016tests}. 

The physical properties of the boundary are also very important in this study. The isothermality of the bounding end-blocks is typically characterized by the Biot number,
\begin{equation}
    Bi = \frac{Nu \, (\lambda/H) }{\lambda_{s}/D_s} \, ,
\end{equation}
where \blue{$\lambda_{s}$} and $D_{s}$ are the solid end-block's thermal conductivity and thickness, respectively. This parameter estimates the effective thermal conductance of the convective fluid layer to that of the solid bounding block.  When $Bi \ll 1$, it is typically argued that boundary conditions are nearly isothermal, since the thermal conductance in the solid so greatly exceeds that of the fluid. We estimate $Bi=0.07$ for the top copper lid and $Bi=0.22$ for the bottom Cu end-block. A similar estimation suggests that $Bi = 0.24$ for both Teflon-coat aluminum boundaries. These $Bi$ values would suggest that boundary thermal anomalies are approximately 10\% of $\Delta T$ \citep[e.g.,][]{verzicco2004}.  

This estimate, however, is not accurate in moderate $Pe$, low $Pr$ liquid metal convection \citep{vogt2018jump}, where the convective flux is predominantly carried by large-scale inertial flows with thermal anomalies that approach $\Delta T$.  These large amplitude thermal anomalies tend to generate significant signals on the container boundaries.   

Furthermore, in low to moderate $Pe$ liquid metal convection, higher $Nu$ implies larger interior temperature gradients since the convective heat flux is carried by large-scale, large amplitude temperature anomalies, instead of via small-scale turbulent plumes \citep[e.g.,][]{grossmann2004}. These temperature anomalies imprint on the top and bottom boundaries and create non-isothermal interfacial conditions. We infer from our $T_{ij}^k$ data that significant interfacial non-isothermality exists in our experiments and that these interfacial thermal anomalies can generate thermoelectric currents that drive long-period dynamics in our TE-MC cases at $0.1 \lesssim N_{\mathcal{C}} \lesssim 1$.

%% file: sections/4a_MCResults.tex
\label{sec:insulated}
\begin{figure}
    \centering
    \makebox[\textwidth][c] {\includegraphics[width=\columnwidth]{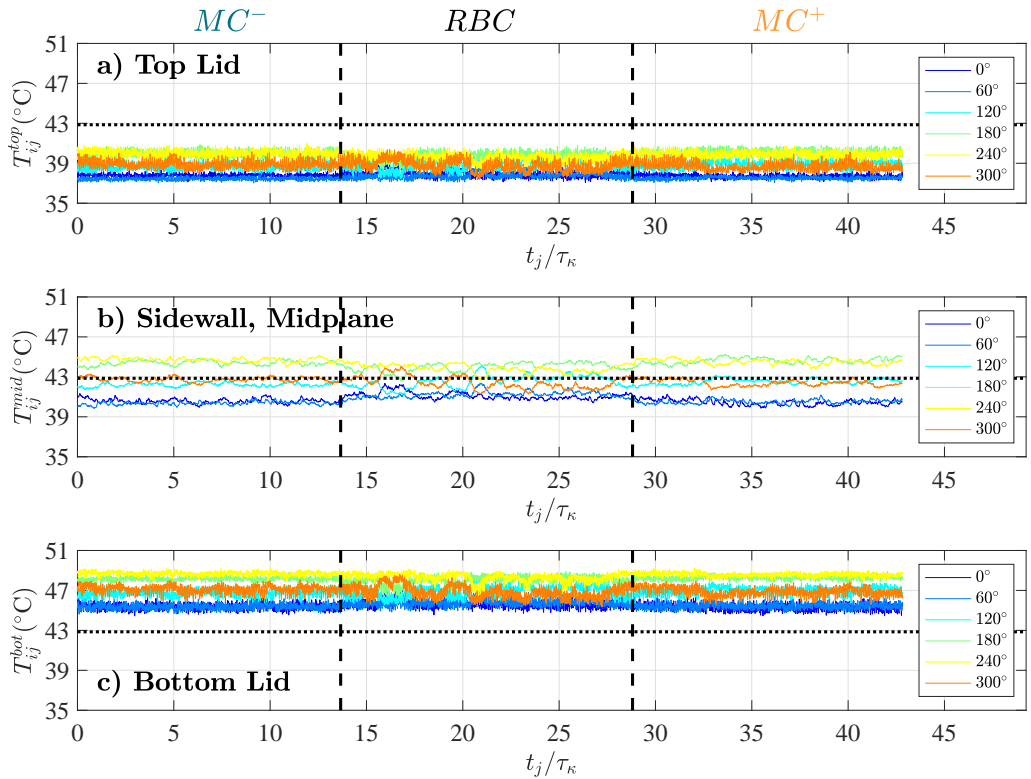}}
  	  	\caption{Temperature time series for the $Ra = 1.61\times 10^6$, $Nu = 5.8$ electrically-insulated (Teflon) boundary conditions experiment.  Data from thermistors, with locations shown in Figure \ref{fig:apparatus}d, embedded in the top boundary $T_{ij}^{top}$ in (a); located on the exterior of the acrylic sidewall mid-plane $T_{ij}^{mid}$ in (b); and embedded in the bottom boundary $T_{ij}^{bot}$ in (c). The mean fluid temperature is $\overline{T} = 42.90 \mathrm{^\circ \, C}$, as marked by the horizontal \blue{dotted} lines in each panel. The abscissa shows the time normalized by the thermal diffusion time scale $t / \tau_\kappa$.  This experiment contains three successive subcases that are divided by two dashed vertical lines: $Insulating\ MC^-$, $Insulating\ RBC$, and $Insulating\ MC^+$.  No significant differences are found between the $Insulating\ MC^-$ and $Insulating\ MC^+$ cases, as is expected for non-thermoelectric, quasistatic magnetoconvection. See Table \ref{app:table1} for detailed parameter values.}
  \label{fig:teflonshorttimeseries}
\end{figure}

A baseline experiment is presented first in which the boundaries are electrically insulating.  Aluminum end-blocks coated in Teflon ($\sigma \approx 10^{-24}\ \mathrm{S/m}$) are used in conjunction with an acrylic sidewall. The Rayleigh number is fixed at $Ra = 1.61\times10^6$ and the equilibrated experiment is run continuously for $t = 42.8 \, \tau_\kappa = 9.6\times 10^3 \, \tau_{f\!f}$.  During this $8.9$-hour data acquisition, three separate sub-experiments are carried out. During the first $13.6 \, \tau_\kappa$, an $120\ \mathrm{gauss}$ downwardly directed ($\bm{\hat e_b} = -\bm{\hat e_z}$) magnetic field is applied, such that $Ch = 2.42\times10^3$, and $N_\mathcal{C} = 0.31$. This sub-case is called $Insulating\ MC^-$. The magnetic field is set to zero in the next sub-case, $Insulating\ RBC$, which extends from $t = 13.6 \, \tau_\kappa$ to $28.8 \, \tau_\kappa$.  The $120\ \mathrm{gauss}$ magnetic field is turned back on, but its direction is flipped such that it is directed upwards ($\bm{\hat e_b} = + \bm{\hat e_z}$) in the last sub-case, $Insulating \ MC^+$, which runs from $t = 28.8 \, \tau_\kappa$ to $42.8 \, \tau_\kappa$.  The Nusselt number  is approximately constant, $Nu \simeq 5.8$, in all three sub-cases.  (See Table \ref{app:table1} for detailed parameter values.)

\begin{figure}
  \centering
  \makebox[\textwidth][c]
  	{\includegraphics[width=1.05\textwidth]{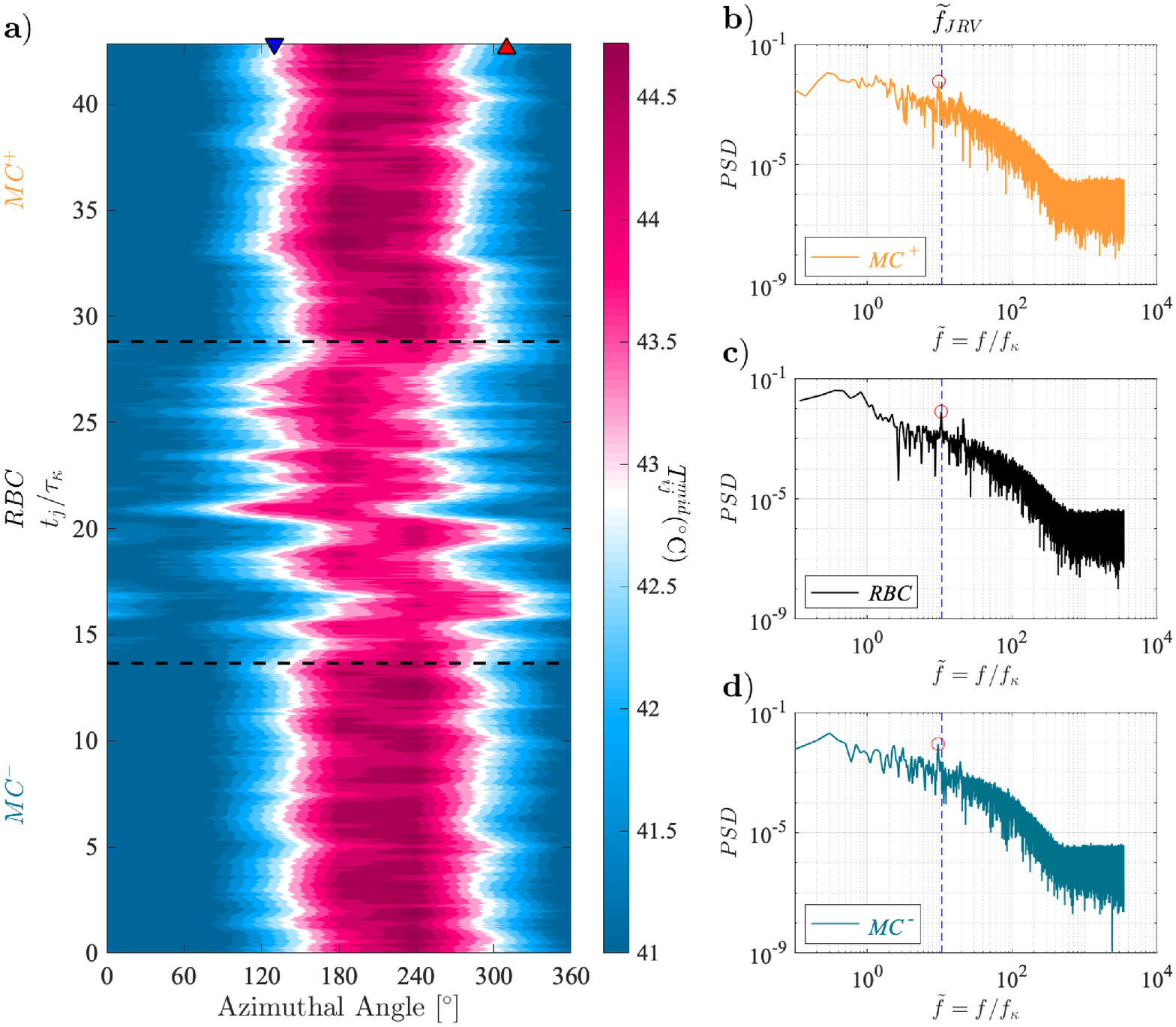}}
  	\caption{Electrically-insulating boundary study: (a) a contour map of the mid-plane sidewall temperature field $T_{ij}^{mid}$ in the $Ra = 1.61\times 10^6$ case \JA{(i.e., corresponding to fig. \ref{fig:teflonshorttimeseries}(b))}. The horizontal axis shows the azimuthal angle around the tank; the vertical axis shows time normalized by $\tau_\kappa$. The blue, downward (red, upward) triangle on the top axis denotes the azimuth of the heat exchanger inlet (outlet) location. The black dashed lines separate the $Insulating\ MC^-$, $Insulating\ RBC$ and $Insulating\ MC^+$ subcases. Hann-windowed \JA{FFTs} of the temperature data from the midplane thermistor located at $120^\circ$ are shown for (b) the $Insulating\ MC^+$ subcase; (c) the $Insulating\ RBC$ subcase; (d) the $Insulating\ MC^-$ subcase. The red circles mark the lowest frequency sharp spectral peaks that correspond to the empirical characteristic frequency prediction for turbulent RBC, \citep{vogt2018jump} $\widetilde f_{JRV} = f_{JRV}/f_\kappa \approx 10.77$, shown as the blue dashed vertical lines in each spectrum. In the $Insulating\ RBC$ case, the distinct sharp peak frequency normalized by the thermal diffusion frequency $f_{peak}/f_\kappa \approx 10.51$. This agrees within $2.5\%$ with $\widetilde f_{JRV}$.}
  \label{fig:teflonbflip}
\end{figure}

Figure \ref{fig:teflonshorttimeseries} shows the temperature time series from the electrically-insulating experiment on a) the top end block $T_{ij}^{top}$, (b) the sidewall midplane $T_{ij}^{mid}$, and (c) the bottom end block $T_{ij}^{bot}$. The horizontal axis shows time normalized by the nondimensional thermal diffusion time $t/\tau_\kappa$. In each panel, the line color represents an individual thermistor, each spaced $60$ degrees apart in each layer (as shown in Figure \ref{fig:apparatus}).

The temperature time series in the midplane \blue{contains less high-frequency} variance relative to the top and bottom block thermistor signals because the measurement \blue{is} taken outside the acrylic sidewall, and thus \blue{is} damped by skin effects. The temperatures in the top block are all well below the mean temperature of the fluid \blue{(black dotted line)}; the midplane temperatures are adequately situated around the mean temperature line, and the bottom block temperatures are all well above the mean temperature.  However, the temperature range in each panel covers nearly 50\% of the mean temperature difference $\Delta T$ across the fluid layer.  This implies strong horizontal temperature anomalies exist in the end blocks, even though the Biot numbers for this experiment is well below unity ($Bi \simeq 0.24$). The RBC case features slightly lower peak-to-peak temperature variations in the midplane thermistors, along with a slightly higher variance in each time series. This suggests that the $RBC$ case carries more of the convective heat flux via higher speed, magnetically undamped flows with regards to the $MC^-$ and $MC^+$ cases. Importantly for later comparisons to cases with electrically-conducting boundaries, the $MC^-$ and $MC^+$ \blue{cases} are essentially identical in all their statistical properties and behaviors. Thus, these two MC cases are not sensitive to the direction of $\boldsymbol B$, as is expected in quasi-static, non-thermoelectrically-active magnetoconvection. 

Figure \ref{fig:teflonbflip}(a) shows the spatiotemporal evolution of the midplane temperature data $T_{ij}^{mid}$ in the electrically insulating experiment. The colormap represents the temperature, in which red (blue) regions are hotter (colder) relative to the mean value (white). The midplane temperature field contains a warmer region on one side of the tank and a downwelling region antipodal to that, as \JA{found in RBC cases with a single LSC} \cite[e.g.,][]{brown2007large, vogt2018jump, zurner2019combined}. Thus, we argue based on figure \ref{fig:teflonbflip}(a) that a turbulent LSC is present in these electrically-insulating boundaries $N_\mathcal{C} \lesssim 1$ experiments, and that it maintains a nearly fixed azimuthal alignment for over $40 \, \tau_\kappa$.

\JA{Figures \ref{fig:teflonbflip}(b)-(d) show the} spectral power density of the averaged temperature signals from each horizontal plane plotted versus normalized frequency, $\tilde{f} = f/f_\kappa$. The vertical dashed lines denote the normalized frequency predictions, $\tilde{f}_{JRV}$, for the jump rope LSC described in \citet{vogt2018jump}. The lowest frequency sharp spectral peaks correspond to the JRV frequency \JA{and} are marked with red circles, matching that of \cite{vogt2018jump} to within $2.5\%$ in the $Insulating\ RBC$ case. (The broad lower frequency peaks correspond to the slow meanderings of the LSC plane.) \JA{The distinct sharp peaks in both the $Insulating\ MC^+$ and the $Insulating\ MC^-$ FFTs are $\approx 25\%$ lower than $\tilde{f}_{JRV}$.} We infer then, based on figure \ref{fig:teflonbflip}, that a quasi-stationary turbulent LSC flow is maintained in these electrically insulating, $N_\mathcal{C} < 1$ experiments. The magnetic field does, however, cause a roughly $25\%$ decrease in the LSC oscillation frequency, likely because magnetic drag reduces the characteristic flow speeds. This agrees adequately with eq. (\ref{equ:Zurner}), which predicts a $20\%$ decrease in flow speed at $N_\mathcal{C} = 0.31$.

Following prior LSC studies \blue{\citep[e.g.,][]{cioni1997strongly, brown2009origin,xi2009origin,zhou2009oscillations}}, we approximate the horizontal temperature profile as a sinusoid varying with azimuth angle $\phi$ at each point in the time series: 
\begin{equation}
    T_{fit}^k(t_j)= A^k_j \cos \left(\phi - \xi_j^k \right) + T^k_j .
\label{equ:Tfit}
\end{equation}
On each $z$-level, $A^k_j$ is the instantaneous amplitude of the sinusoidal temperature variation, $\xi_j^k$ denotes the instantaneous azimuthal orientation of the LSC plane, and the instantaneous azimuthal-mean temperature is $T^k_j$.  Using eq. (\ref{equ:Tfit}), we best fit each $z$-level's temperature data at every time step. 
\begin{figure}
  \centering
    \includegraphics[width = \textwidth]{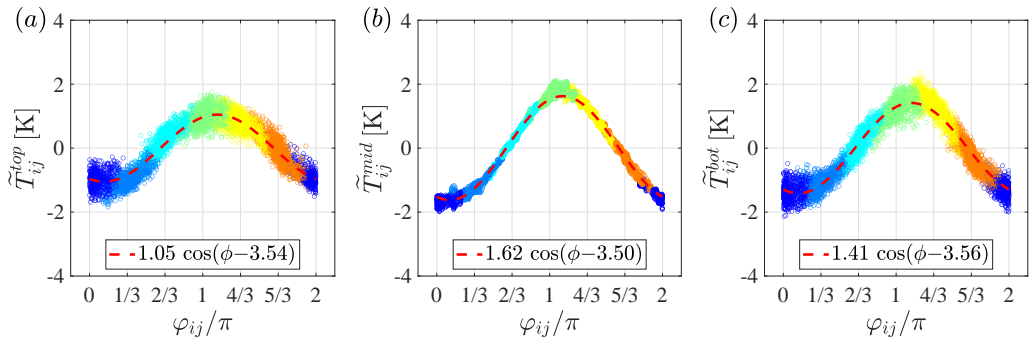}
    \caption{Temperature data of the $Insulating\ RBC$ case shifted azimuthally into the best fit LSC frame, $\widetilde{T}_{ij}^k$, in (a) the top block; (b) the sidewall midplane; and (c) the bottom block. The vertical axis is the temperature minus the azimuthal mean temperature at each time step. Different colors are used to label the location of the thermistors in the lab frames as the LSC \blue{fluctuates} around its mean position \JA{following the color scale convention used in figure \ref{fig:teflonshorttimeseries}. The colors from left (blue) to right (orange) correspond to} thermistors $i = 1$ to $6$, respectively. The time-averaged best fit sinusoidal temperature profile is shown via the dashed red line in each panel.}
    \label{fig:cosRBC}
\end{figure}

Figure \ref{fig:cosRBC} shows the {\it insulating RBC} temperature anomaly on the top plane (a), midplane (b), and bottom planes (c), but with the data at each time step azimuthally-shifted into the best-fit LSC frame. This is accomplished by plotting $\widetilde{T}^k_{ij}$, defined as  
\begin{equation}
\widetilde{T}^k_{ij} \equiv  T_{ij}^k(\varphi_{ij}^k, t_j) - T_j^k  \quad \mbox{where} \quad     \varphi^k_{ij} \equiv \phi_{ij}^k - \left[ \xi_j^k - \xi \right] \, .
    \label{equ:TildeTijk}
\end{equation}
The new azimuth variable $\varphi^k_{ij}$ shifts each instantaneous thermistor measurement $T_{ij}^k$ to its azimuthal location relative to the best fit LSC azimuthal orientation angle $\xi_j^k$ in eq. (\ref{equ:Tfit}). The best fit LSC orientation angle averaged over time and over $z$-level is $\xi = 3.55$ rad for this case. The time-mean best fit sinusoid for the data on each $z$-level is plotted as a dashed red in each panel, with the best fit given in the legend box. \JA{The color of each thermistor follows the convention used in figure \ref{fig:teflonshorttimeseries}.}  
The well-defined patches of color in figure \ref{fig:cosRBC} are aligned with the individual thermistor locations, producing a rainbow color pattern. The relative fixity of these color patches shows that  the approximately sinusoidal temperature pattern does not drift significantly in time in this sub-case. Although they are not shown here, similar rainbow patterns also exist for the two insulating MC sub-cases.  

In sum, we take figures \ref{fig:teflonshorttimeseries} through \ref{fig:cosRBC} as evidence of a quasi-stationary, container-scale LSC in all three electrically insulating sub-cases made with $\Nc \lesssim 0.3$.

%% file: sections/4b_TEMCResults.tex
\label{sec:conducting}

\begin{figure}
    \centering
    \makebox[\textwidth][c] {\includegraphics[width=\columnwidth]{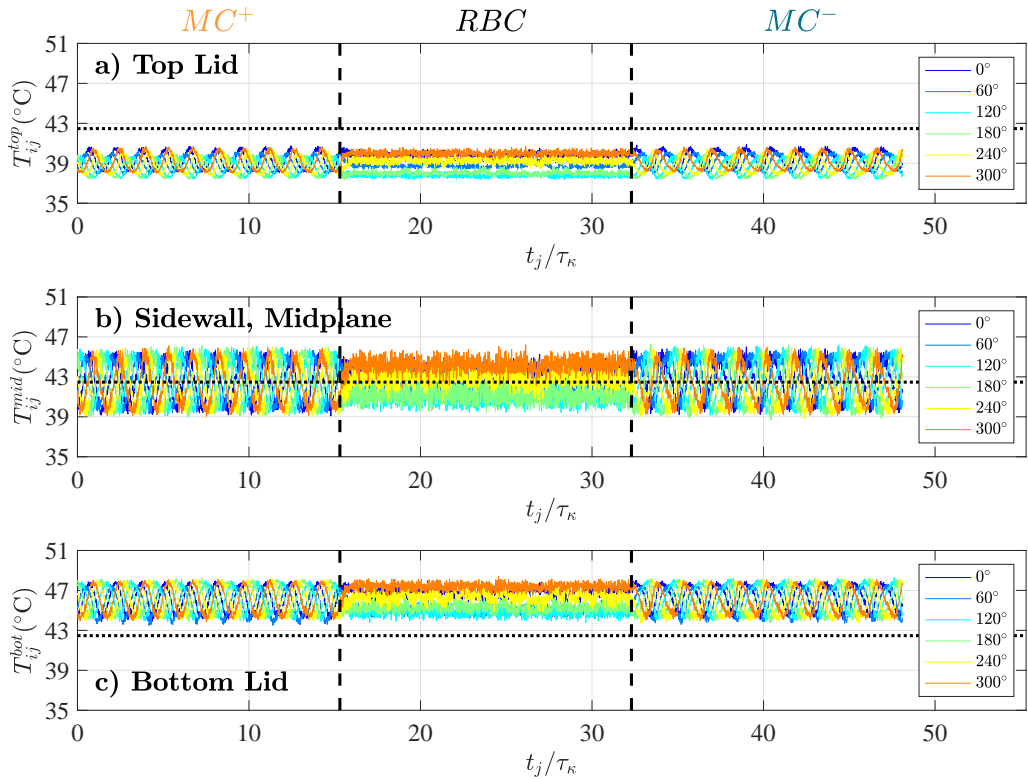}}
      	  	\caption{Temperature time series for the $Ra \simeq 1.82\times 10^6$, $Nu \simeq 5.86$ electrically-conducting boundary conditions experiment.  Data from thermistors, with locations shown in Figure \ref{fig:apparatus}d, embedded in the top boundary $T_{ij}^{top}$ in (a); located on the exterior of the acrylic sidewall mid-plane $T_{ij}^{mid}$ in (b); and embedded in the bottom boundary $T_{ij}^{bot}$ in (c). The mean fluid temperature is $\overline{T} = 42.47 \mathrm{^\circ \, C}$, as marked by the horizontal \blue{dotted} lines in each panel. The abscissa shows the time normalized by the thermal diffusion time scale $t / \tau_\kappa$.  This experiment contains three successive subcases that are divided by two dashed vertical lines: $Conducting\ MC^-$, $Conducting\ RBC$ and $Conducting\ MC^+$ (Table \ref{app:table1}).  Large amplitude, low frequency thermal oscillations are observed at all thermistor locations in the $Conducting\ MC^+$ and $Conducting\ MC^-$ subcases, which differs greatly with respect to the corresponding $Conducting\ MC$ subcases in figure \ref{fig:teflonshorttimeseries}.}
  \label{fig:shorttimeseries}
\end{figure}

\begin{figure}
  \centering
  \makebox[\textwidth][c]
  	{\includegraphics[width=1.05\textwidth]{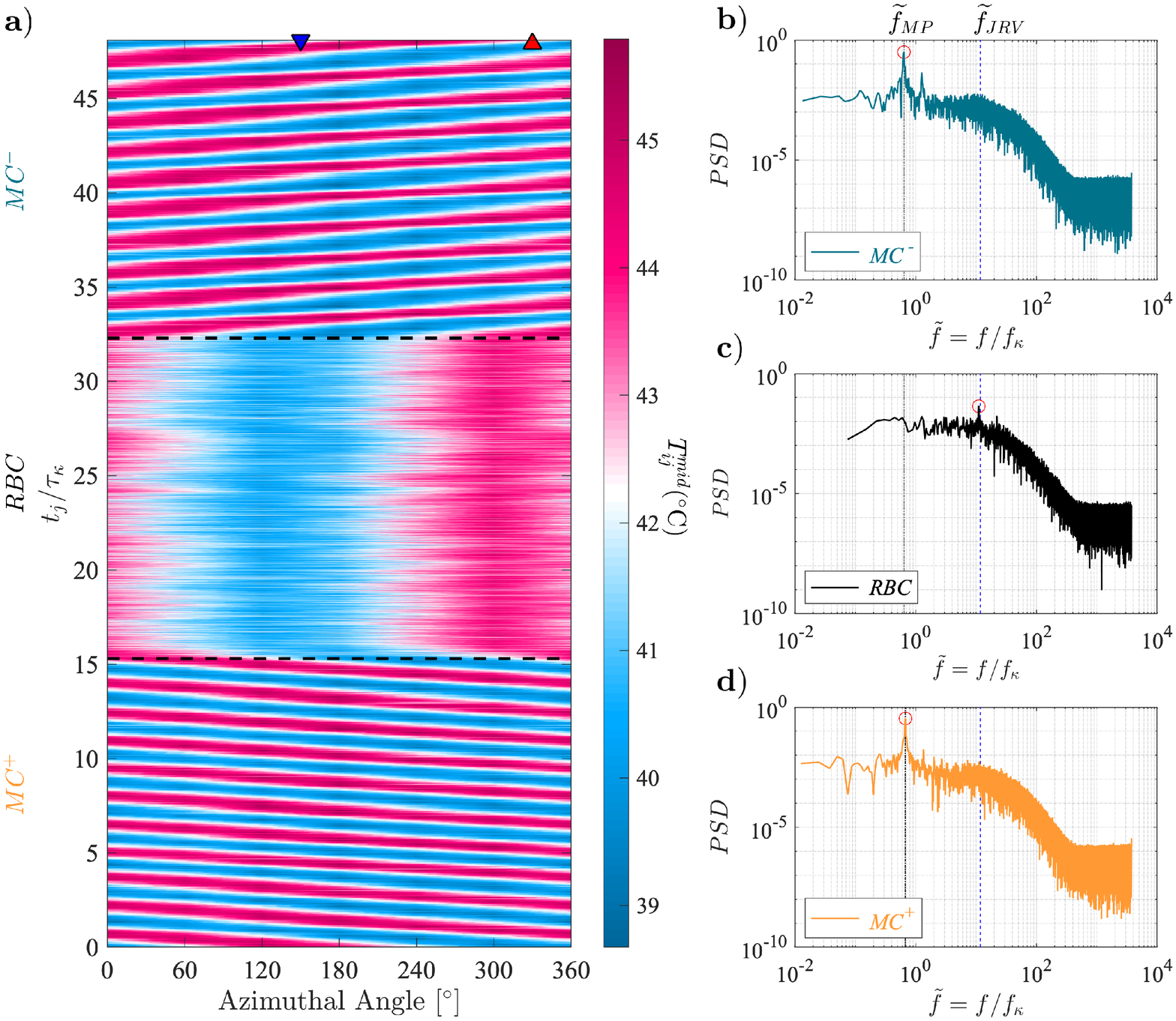}}
\caption{Identical to figure \ref{fig:teflonbflip}, but showing the $Ra \approx 1.8\times 10^6$, $Nu \approx 5.8$ $Conducting\ MC^+$, $Conducting\ RBC$, $Conducting\ MC^-$ subcases experiment.  However, all the FFTs here are analyzed using the $Long$ version of the same experiments shown in appendix table \ref{app:table1}. The averaged low frequency spectral peak in the $Conducting\ MC$ subcases is marked by the vertical black dot-dashed lines in (b), (c) and (d). This corresponds to the magnetoprecessional (MP) mode and its nondimensional frequency is labeled $\widetilde f_{M\!P}$. }
  \label{fig:bflip}
\end{figure}

\begin{figure}
  \centering
    \includegraphics[width = \textwidth]{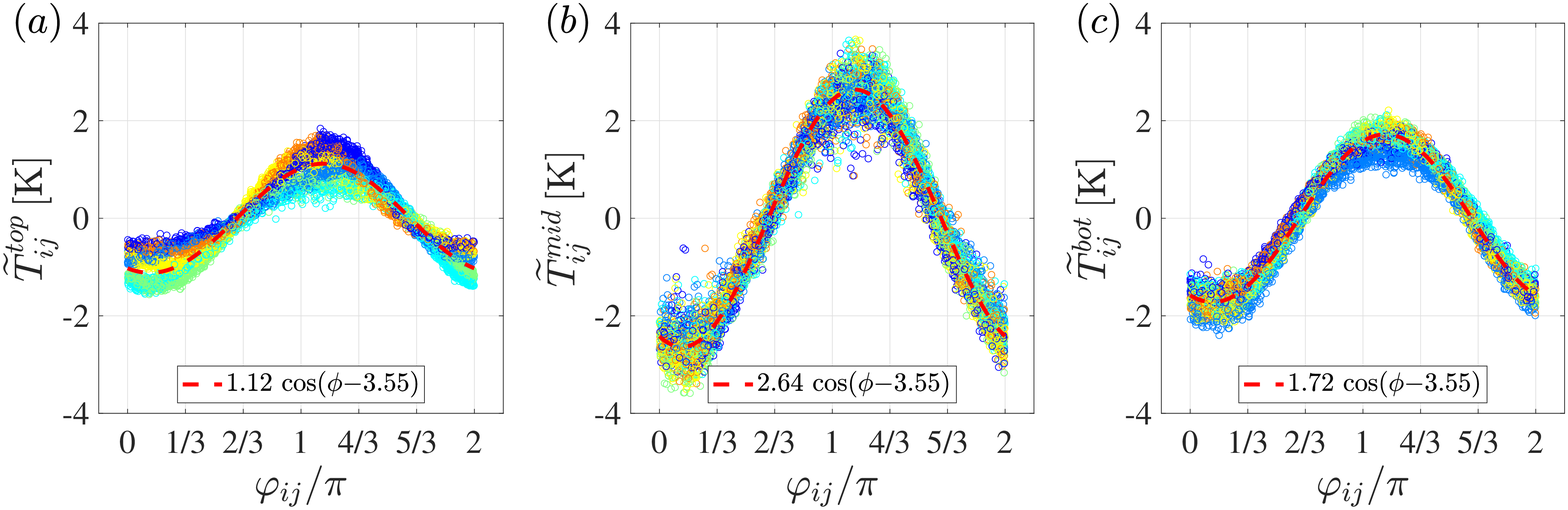}
    \caption{Temperature anomaly $\widetilde{T}_{ij}^k$ as defined eq. (\ref{equ:TildeTijk}) on the a) top; b) midplane; and c) bottom horizontal planes in the LSC frame of the $Conducting\ MC^-$ subcase. For ease of comparison, we set $\xi = 3.55$, which is the same as the $Insulating\ RBC$ case, since the precessing case does not have a meaningful time-averaged LSC position. The same colors are also used to label the location of the thermistors in the lab frames as figure \ref{fig:cosRBC}. Contrary to figure \ref{fig:cosRBC}, where the same color data cluster near a fixed azimuth, here each color is spread out and covers the entire azimuth relative to the LSC plane, which occurs because the LSC plane is constantly precessing through all the azimuthal angles. Panels (a) -- (c) show that a sinusoidal temperature profile exists at each horizontal level $k$, with the largest amplitude in the midplane. \JA{The time-averaged best fit sinusoidal temperature profile is shown via the dashed red line in each panel.}}
    \label{fig:cosMP}
\end{figure}

Another approximately $48 \tau_\kappa$ experiment has been carried out, but with all the boundaries electrically conducting such that thermoelectric effects can now affect the system, in contrast to the electrically insulating experiment presented in \S \ref{sec:insulated}.  The end-blocks used here are copper and the sidewall is stainless steel 316L. The heating power is fixed at 396.2W, leading to $Ra \simeq 1.8 \times 10^6$ and $Nu \simeq 5.9$.  The experiment is made up of three successive sub-cases, $Conducting\ MC^+$, $Conducting\ RBC$, and $Conducting\ MC^-$, having an upward 120 gauss applied magnetic field $+\bm{\hat e_z}$, no magnetic field, and then a downward 120 gauss magnetic field $-\bm{\hat e_z}$, respectively. This corresponds to $\Nc \simeq 0.3$ in the two MC sub-cases and $\Nc = 0$ in the RBC case, similar to the prior insulating sub-cases. Figure \ref{fig:shorttimeseries} - \ref{fig:cosMP} correspond to figure \ref{fig:teflonshorttimeseries} - \ref{fig:cosRBC}. The exact parameters are given in table \ref{app:table1} in Appendix \ref{AppC}. 

Figure \ref{fig:shorttimeseries} presents the $T_{ij}^k$ thermistor time series data from this electrically conducting experiment, following the same plotting conventions as figure \ref{fig:teflonshorttimeseries}. The time series \blue{shows} that the $Conducting\ RBC$ sub-case generates a nearly stationary LSC structure, similar to the time series in figure \ref{fig:teflonshorttimeseries}. \blue{In the $Conducting\ MC^+$ and $Conducting\ MC^-$ cases, the temperature signals oscillate periodically around the mean temperature of the same layer height. Moreover, the oscillation of different heights are in phase at each azimuthal position. The temperature measurements indicate} the presence of a container scale, coherent thermal structure that precesses in time. 

%

Figure \ref{fig:bflip} follows the same plotting conventions as figure \ref{fig:teflonbflip}. Figure \ref{fig:bflip}(a) shows a temperature contour map of the midplane sidewall thermistors, $T_{ij}^{mid}$ for the electrically-conducting experiment. In the $Conducting\ RBC$ case, the temperature pattern remains roughly fixed in place, similar to the insulating case. In contrast to this, the temperature field is found to coherently translate in the $-\bm{\hat e_\phi}$ direction in the $Conducting\ MC^+$ sub-case and to translate in the $+\bm{\hat e_\phi}$ direction in the $Conducting\ MC^-$ sub-case.  However, at any instant in time, $t_j$, the azimuthal temperature pattern is similar to that of the LSC-like pattern found in the electrically-insulating experiment, with one warmer region and an antipodal cooler region.

Comparing figures \ref{fig:teflonbflip}(a) and \ref{fig:bflip}(a) shows that the instantaneous LSC-like temperature pattern precesses around the container only in MC cases with electrically conducting boundaries. Further, the precession direction depends on the sign of the magnetic field, as cannot be the case for standard quasi-static MHD processes.  Thus, we hypothesize that an LSC exists in these electrically-conducting sub-cases, and that thermoelectric current loops exist across the container's electrically conducting boundaries which drive the LSC to precess azimuthally in time.  Our model for this thermoelectric magnetoprecession (MP) process is presented in \S \ref{sec:thermoelectric}. 


\JA{The figure \ref{fig:bflip}(a) contour map also reveals that a slight asymmetry exists in the $Conducting \ MC^-$ precession rate that does not exist in the $Conducting \ MC^+$ case. The precessional banding of the temperature field is uniform in the $Conducting MC^+$ case. In contrast, the bands have a slight variation in thickness in the $Conducting MC^-$ case.  We do not currently have an explanation for this difference between the $MC^-$ and $MC^+$ cases.}

Figures \ref{fig:bflip}(b)-(d) show the time-averaged, thermal spectral power density plotted versus normalized frequency for the $Conducting\ MC^+$, $Conducting\ RBC$, and $Conducting\ MC^-$ sub-cases. To better identify the spectral peaks, these FFTs are made using three longer experimental cases, each up to $\approx 100 \tau_\kappa$ in duration but employing the same control parameters. (Detailed parameter values are provided in Table \ref{app:table1}.) The frequencies are normalized by the thermal diffusion frequency $\widetilde{f} = f/f_\kappa = f  \tau_{\kappa}$. Red circles mark the peak frequency in each spectrum. The peak of the $Conducting\ RBC$ sub-case is in good agreement with the predicted jump rope vortex frequency \JA{$\widetilde f_{JRV} = 11.79$ } (dashed blue vertical line). The magnetoprecessional frequency dominates the $Conducting\ MC^-$ and $Conducting\ MC^+$ spectra in figures \ref{fig:bflip} (b) and (d), respectively.  \JA{The peak frequencies are nearly identical in $Conducting\ MC^-$ and $Conducting\ MC^+$ cases, with a mean value \JA{$\widetilde f_{M\!P}=0.66$ } (black dot-dashed vertical line). Thus, magnetoprecession is slow relative to the jump rope mode, with $\widetilde f_{M\!P} = 0.06 \widetilde f_{JRV}$.} 

Figure \ref{fig:cosMP} is constructed parallel to figure \ref{fig:cosRBC}, but plots the horizontal temperature anomalies of the $Conducting\ MC^-$ thermistor data azimuthally-shifted into the best fit LSC reference frame. Since the LSC continually precesses in the $+\bm{\hat e_\phi}$ direction in this sub-case, there is no mean location of the best fit LSC plane.  For ease of comparison with figure \ref{fig:cosRBC}, we set $\xi= 3.55$. In figure \ref{fig:cosRBC}, each thermistor's data exists in an azimuthally-localized cloud since the LSC maintains its position over time.  In contrast,  each thermistor's data points form an approximately continuous sinusoid in this magnetoprecessional case. This occurs since the thermal field precesses past each of the spatially fixed thermistors and, thus, each thermistor samples every part of the sinusoidally precessing temperature field over time. 

The top block thermistor data sets in figure \ref{fig:cosMP}(a) deviate from that of a sinusoid. This is caused by spatially fixed $\simeq 0.5\, \mathrm K$ temperature anomalies in the top block that are co-located with the inlet and outlet positions of the top block heat exchanger's cooling loop, which are located at $\phi = 150^{\circ}$ and $330^{\circ}$, respectively. These fixed temperature anomalies are likely not evident in figure \ref{fig:cosRBC} because the orientation angle of the LSC remains nearly aligned with the heat exchanger inlet and outlet angles in the electrically-insulating experiment. In addition, we note that the midplane has a larger temperature variation in figure \ref{fig:cosMP}(b) than in the corresponding $Insulating \ RBC$ case. This may be due to differences in $Bi$ for the differing experiments.

\subsection{Fixed \texorpdfstring{$Ra \approx 2\times 10^6$}\ TEMC Survey}
%
\begin{figure}%
    \centering
    \includegraphics[width =\textwidth]{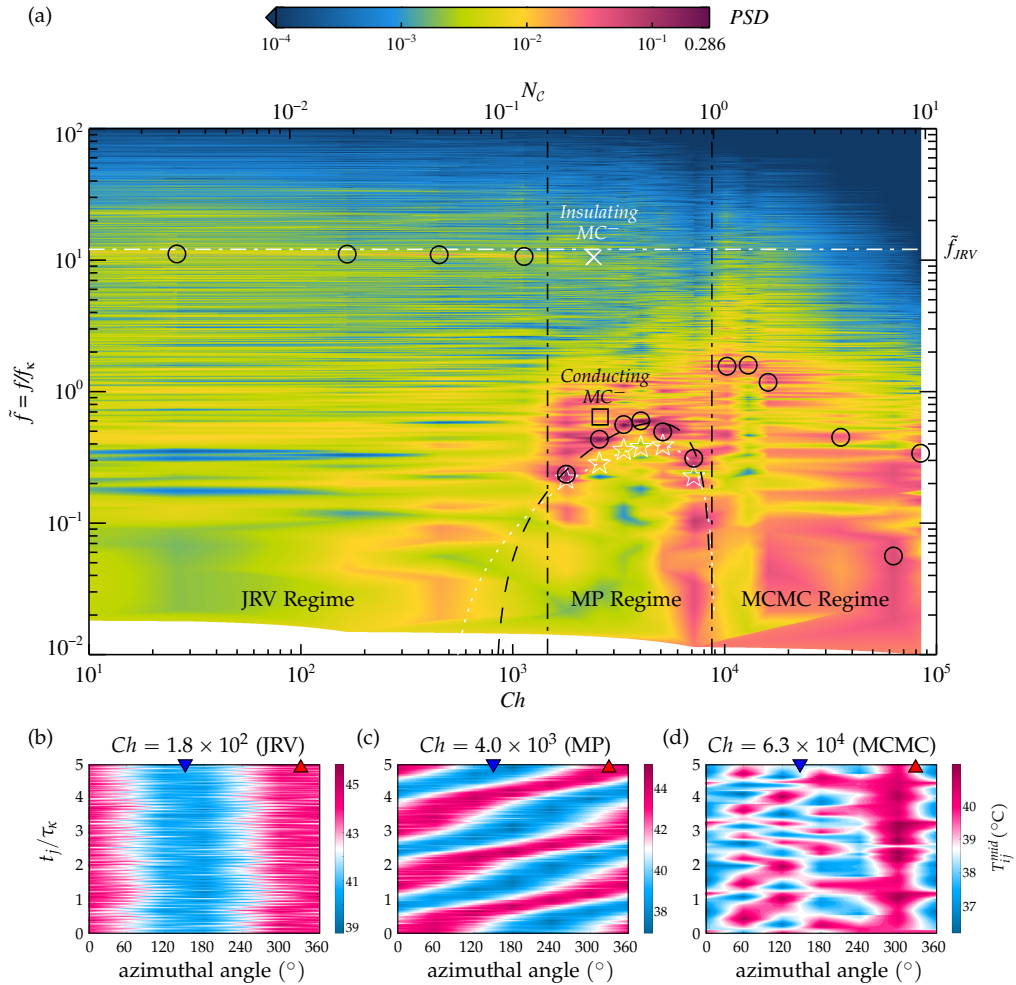}
    \caption[Spectrogram]{(a) \blue{Power spectral density (PSD) of the $T_{ij}^{mid}$ temperature data versus $Ch$ and $N_{\mathcal C}$ for the $Ra \approx 2\times 10^6$ cases shown in table \ref{app:table2}. The vertical axis is thermal diffusion frequency $\widetilde f = f/f_\kappa$.
    The peak frequency for each case is marked with black open circles. The interaction parameter is calculated here as $ N_\mathcal{C} = \sqrt{Ch^2 Pr/Ra_0}$, where $Ra_0 = 2.12 \times 10^6$ 
 \JA{corresponds to}
 the case with no magnetic field. The jump rope vortex (JRV), the magnetoprecession (MP), and the multi-cellular magnetoconvection (MCMC) regimes are separated by the two black vertical dot-dashed lines. 
    The JRV frequency \citep{vogt2018jump}, $\widetilde f_{JRV}$, is shown as the white horizontal dotted line near $\widetilde f \approx 12.1$. The black square marks the peak frequency of the $Conducting\ MC^-$ case, and the white cross marks the peak frequency of the $Insulating\ MC^-$ case. 
    The black dashed curve denotes the second-order fit to the experimental data in the MP regime.
    The white stars are magnetoprecession frequency estimates calculated using eq.~(\ref{equ:pmtheory}) for each MP case, and the white dotted curve is the second-order fit of these theoretical estimates developed in \S \ref{sec:exp_veri}. The lower panels show sidewall midplane temperature contour maps in (b) JRV, (c) MP, and (d) MCMC regimes. The blue downwards and red upwards triangles in the lower panels denote the heat exchanger inlet and outlet azimuth locations, respectively.}}
    \label{fig:spg}%
\end{figure}

To characterize the system's behavioral regimes, we have conducted a survey of turbulent TEMC with the Rayleigh number fixed at approximately $2\times 10^6$ and the Chandrasekhar number varying from $0$ to $8\times 10^5$ \blue{all with a vertically downward applied magnetic field ($\bm{\hat e_b} = -\bm{\hat e_z}$)}. Three regimes are found: (i) the jump rope vortex ($JRV$) regime; (ii) the magnetoprecessional ($MP$) regime; and (iii) the multi-cellular magnetoconvection ($MCMC$) regime. 

Figure \ref{fig:spg}(a) shows a thermal spectrogram made using the $T_{ij}^{mid}$ data, plotted as functions of  $f /f_\kappa$ on the vertical axis and the Chandrasekhar number $Ch$ on the horizontal axis. Here, we use $Pr = 0.027$, and the RBC case's $Ra_0 = 2.12 \times 10^6$ value to calculate $N_\mathcal{C}$ for all the cases. The peak frequency at each $Ch$-value is marked by an open black circle. Starting from the left of the figure, the predicted peak RBC frequency, $\tilde{f}_{JRV}$ derived from equation eq. (\ref{equ:jrv}), is marked by the white, horizontal dashed line. In the JRV regime, the $0 < N_\mathcal{C} \lesssim 10^{-1}$ experimental data are in good agreement with $\tilde{f}_{JRV}$, with sidewall thermal fields that correspond to that of an LSC-like flow (e.g., figure \ref{fig:spg}(b)). In this regime, buoyancy-driven inertia is the dominant forcing in the system. As $N_\mathcal{C}$ increases and exceeds unity, it is expected that the LSC will weaken and eventually disappear \citep[e.g.,][]{cioni2000, zurner2020flow}. However, the MP regime exists in the intermediate $N_\mathcal{C}$ TEMC system. In this regime, the spectral peak switches from near to $\tilde{f}_{JRV}$ to the slow magnetoprecessional frequency above $N_\mathcal{C} \approx 0.1$, corresponding to the magnetoprecessional sidewall thermal signal shown in figure \ref{fig:spg}(c)). The MP frequency grows with $N_\mathcal{C}$, reaching a value of $0.60\ f_\kappa$ near $N_\mathcal{C} \approx 0.4$. At higher $N_\mathcal{C}$, the peak frequency decays, becomes unstable, and mixes with other complex modes at $N_\mathcal{C} \gtrsim 1$ (e.g., figure \ref{fig:spg}(d)). The single, turbulent LSC likely gives way to multi-cellular bulk flow in this $N_\mathcal{C} \gtrsim 1$ MCMC regime. 

We contend that it is the existence of coherent thermoelectric current loops existing across the top and bottom horizontal interfaces of the fluid layer that drive the magnetoprecessional mode observed in the MP regime. Following the arguments of \S \ref{sec:TEeffects}, this requires horizontal temperature gradients to exist along on these bounding interfaces as shown schematically in figure \ref{fig:TE_AB}.  To quantify this, 
the horizontal temperature difference at height $z_k$ and time $t_j$ is estimated using the best fit of the data to eq. (\ref{equ:Tfit}) as 
\begin{equation}
    \delta T_j^k=  \mathrm{max} \left( T_{fit}^k(t_j)\right) - \mathrm{min} \left(T_{fit}^k(t_j)\right)   =  2 A_j^k \, .
    \label{equ:deltaT_h1}
\end{equation}
Its time-mean value is denoted by 
\begin{equation}
    \delta T^k  = \frac{2}{N} \sum\limits_{j=1}^{N} A_j^k \, , 
    \label{equ:deltaT_h2}
\end{equation}
where $j$ is the $j^{\mathrm{th}}$ step in the discrete temperature time series and $N$ is the total number of time steps. Thus, $\delta T^{top}$ estimates the time-averaged, maximum horizontal temperature difference in the top block thermistors located at $z = 100.6\ \mathrm{mm} = 1.022 H$ and $r \blue{=0.71} R$. Similarly, $\delta T^{bot}$ estimates this value using the bottom block thermistors located at $z = -2.0 \ \mathrm{mm} = -0.020 H$ and $r \blue{=0.71} R$. 

\begin{figure}
    \centering
    \includegraphics[width=\textwidth]{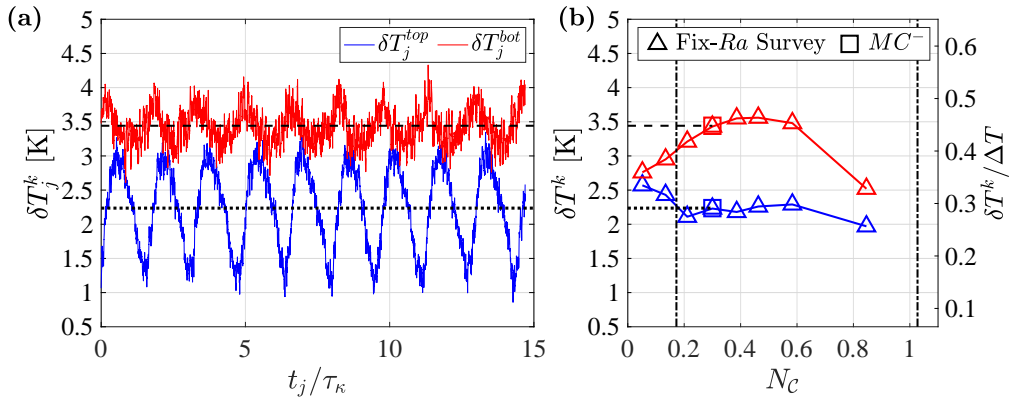}
    \caption{(a) \blue{Time series of the horizontal temperature difference \JA{at} different heights} $\delta T_j^k$, defined in eq. (\ref{equ:deltaT_h1}), from the $Conducting\ MC^-$ case at $Ra = 1.83\times 10^6$, $Ch = 2.59\times10^3$, and $N_\mathcal{C}= 0.31$. The horizontal axis is normalized time $t_j/\tau_\kappa$. \blue{The black dotted line denotes the mean values of $\delta T^{top} = 2.24\ \mathrm K$, and the black dashed line denotes} $\delta T^{bot} = 3.44\ \mathrm K$. (b) Time-averaged horizontal temperature difference estimates $\delta T^{k} $ on the top and bottom boundaries for the fixed-$Ra$ cases \blue{($Ra\approx 2\times 10^6$) at $N_{\mathcal{C}} < 1$. The fixed-$Ra$ cases are marked by triangles; blue (red) color represents top (bottom) boundary measurements. The magnetoprecessional (MP) regime lies between the two vertical dot-dashed lines.} The right hand $y$-axis denotes $\delta T^{k}$ normalized by \blue{the averaged vertical temperature difference $\Delta T = 7.68\ \mathrm{K}$ of the fixed-$Ra$ cases shown here. Values of $\delta T_j^k$ for the $Conducting\ MC^-$ case are marked by the square symbols.}}
    \label{fig:delta_deltaT}
\end{figure}

Figure \ref{fig:delta_deltaT}(a) shows time series of the maximum \JA{\textit{horizontal}} temperature variations in the top and bottom boundaries in the reference \JA{frame of the fitted LSC plane, $\delta T_j^k$, calcuated using eq. (\ref{equ:deltaT_h1})}. This data is from our canonical $Conducting\ MC^-$ case at $Ra = 1.83\times 10^6$, $Ch = 2.59\times10^3$, and $N_\mathcal{C}= 0.31$. \blue{The temperature time series for this case has been shown in figure \ref{fig:shorttimeseries}}. 
In the top block, the time-averaged maximum horizontal temperature variation, plotted in blue, is $ 2.24\ \mathrm{K}$, which is $1.2\ \mathrm{K}$ smaller than that of the bottom block (in red), $3.44\ \mathrm K$. Moreover, the top has a larger magnitude of fluctuation of $\sim 2\ \mathrm{K}$, while the bottom remains relatively stable with a fluctuation of $\sim 1\ \mathrm{K}$. \JA{The difference in top and bottom $\delta T_j^k$ fluctuation amplitudes} is likely due to the structure of the heat exchanger, in which the inlet and the outlet of the cooling water are antipodal to one another. This imposes a small \JA{($\sim 1$ K)}, spatially fixed temperature gradient along the line connecting these two points. This causes the horizontal temperature variation at the top boundary to fluctuate as the LSC precesses across the top block's spatially fixed temperature gradient. We hypothesize that this fluctuation propagates to the bottom boundary, generating a smaller fluctuation there than on the top block and lagging the top fluctuation by about $0.6$ thermal diffusion times. 

%

Figure \ref{fig:delta_deltaT}(b) shows $\delta T^k$, the time-mean horizontal temperature differences \JA{calculated via eq. (\ref{equ:deltaT_h2})}on the top block \blue{(blue triangles)} and \JA{on the} bottom block (red \blue{triangles}) for the $N_{\mathcal{C}} < 1$ experiments in the fixed $Ra$ survey \blue{($Ra\approx 2\times 10^6$)}. \blue{The $Conducting\ MC^-$ case in panel (a) corresponds to the square markers on the right in panel (b)}. \blue{ The horizontal dashed and dotted lines show the mean values of \blue{$\delta T^k$} for the $Conducting\ MC^-$ case. The two vertical dot-dashed lines denote the boundaries between the JRV, MP and MCMC regimes. In the lowest $N_{\mathcal{C}}$ case shown in figure \ref{fig:delta_deltaT}(b)}, it is found that $ \delta T^{top} \approx \delta T^{bot} \approx 2.6$ K.  This value is nearly $40\%$ of the vertical temperature gradient across the tank, and is similar to values found for comparable RBC cases.  We argue that this 2.6 K value is \JA{predominantly generated} by the jump rope vortex imprinting its thermal anomalies onto the top and bottom boundary thermistors. 
The values of $\delta T^{bot}$ exceed $\delta T^{top}$ in the MP regime ($0.1 \lesssim N_{\mathcal{C}} \lesssim 1$). For $N_{\mathcal{C}} \gtrsim 1$, the jump rope-style LSC breaks down into multi-cellular flow \citep[e.g.,][]{zurner2020flow} and it is not possible to fit a sinusoidal function of the form (\ref{equ:Tfit}) to the thermistor data in the top and bottom blocks.

The right hand vertical axis in figure \ref{fig:delta_deltaT}(b) shows thermal block temperature differences normalized by the vertical temperature difference, $\delta T^{k}/ \Delta T$.  The fixed $Ra$ survey cases have  $\delta T^{k}/ \Delta T$ values ranging from roughly 0.3 to 0.5, demonstrating that low $Pr$ convective heat transfer occurs via large-scale, large amplitude thermal anomalies that may alter the thermal boundary conditions in finite $Bi$ experiments. Such conditions differ from those typically assumed in theoretical models of low Prandtl number convection \citep[e.g.,][]{CleverBusse1981,Thual1992}. 

The slow magnetoprecessional modes only appear in MC experiments with electrically conducting boundaries for $0.1\lesssim \blue{N_\mathcal C} \lesssim 1$. Strong, coherent horizontal temperature gradients exist along the top and bottom boundaries in these cases, as shown in figure \ref{fig:delta_deltaT}.  This suggests that magnetoprecession is controlled by the material properties of the boundaries and the horizontal temperature gradients on the liquid-solid interfaces. Based on these arguments, we develop a simple model for thermoelectrically-driven magnetoprecession of the LSC in the following section. 


%% file: sections/5_TEModel.tex

\blue{This study presents the first detailed characterization of the large-scale, long-period magnetoprecessional (MP) mode that appears in turbulent MC cases with conducting boundaries.} We hypothesize that the MP mode emerges from an imbalance between the thermoelectric Lorentz forces at the top and bottom boundaries of the fluid layer. This imbalance, which arises due to the differing thermal gradients on the top and bottom boundaries, creates a net torque on the overturning LSC. This net torque causes the LSC to precess like a spinning top. To test this hypothesis, a simple mechanistic model of such a thermoelectrically-driven magnetoprecessing LSC is developed, and is shown to be capable of predicting the essential behaviors in our MP system.

\subsection{Angular Momentum of the LSC Flywheel}

A Cartesian coordinate frame is used in our model of thermoelectrically-driven magnetoprecession of the LSC. This Cartesian frame is fixed in the LSC plane such that $\bm{\hat e_y}$ always points along the $\xi_j = 0$ direction.  The thermal gradient is also aligned in the same direction, yielding $\bm{\hat e_n} = \bm{\hat e_y}$.  The magnetic field direction is oriented in $\pm \bm{\hat e_z}$, and the right-handed normal to the LSC plane is oriented in the $\bm{\hat e_x}$-direction.

\begin{figure}
  \centering
    \includegraphics[width = 0.5\textwidth]{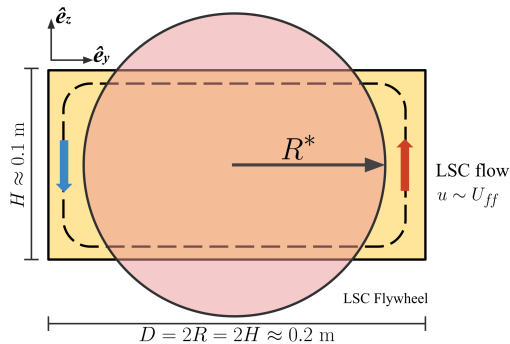}
    \caption{Cross-section view of the precessional flywheel model (pink) in the LSC plane (yellow). The  precessional flywheel is assumed to have the same cross-sectional area as the LSC plane, $\pi (R^*)^2 = 2H^2$. The angular velocity of the overturning LSC flywheel is estimated by assuming it rotates at the free-fall speed $\omega_{LSC} \approx U_{f\!f}/R^*$. }
    \label{fig:Rstar}
\end{figure}

We treat the LSC as a solid cylindrical flywheel that spins around a midplane $\bm{\hat e_x}$-axis, as shown in figure \ref{fig:Rstar}. The LSC is taken to have the same cross-sectional area as the LSC plane, $A_{LSC} = \Gamma H^2$. The corresponding radius of the solid LSC cylinder is then
\begin{equation}
    R^* = \sqrt{A_{LSC}/\pi} = \sqrt{\Gamma H^2/\pi}.
    \label{equ:Rstar}
\end{equation}
which corresponds to $R^* \approx  0.8 H$ for the $\Gamma = 2$ experiments carried out here. 
The volume of \JA{the} turbulent LSC is $V_{LSC}$. For convenience, we take the depth of the LSC in $\bm{\hat e_x}$ to be $R^*$ so that $V_{LSC} = \pi R^{*3}$, noting that the assumed depth and $V_{LSC}$ both drop out of our eventual prediction for the LSC’s magnetoprecession rate $\omega_{M\!P}$. The LSC flywheel, as constructed, does not physically fit within the tank since $R^* > H/2$, as shown in figure \ref{fig:Rstar}. (It is not shown to scale in figure \ref{fig:flywheel}(b).) 

The angular momentum of the flywheel is \JA{taken to be} that of a uniform-density solid cylinder with mass  $M_{LSC} = \rho V_{LSC}$ and radius $R^*$, rotating around $\bm{\hat e_x}$. Its moment of inertia with respect to the $\bm{\hat e_x}$-axis is
\begin{equation}
    I =  \frac{1}{2} M_{LSC} R^{*2} =  \frac{1}{2} \rho V_{LSC} R^{*2} \, .
    \label{equ:moi}
\end{equation}
We use the upper bounding free-fall velocity as an estimate of the angular velocity vector for the LSC flywheel:
\begin{equation}
\bm{\omega}_{LSC} \approx U_{f\!f}/R^* \bm{\hat e_x} \JA{.}
\end{equation}
Thus, the angular momentum due to the overturning of the flywheel, $\bm L_{LSC}$, is oriented along $\bm{\hat e_x}$ and is estimated as: 
\begin{equation}
    \bm L_{LSC} = I \bm \omega_{LSC} \approx \left( \frac{1}{2} \rho V_{LSC} R^{*2} \right) \left( \frac{U_{f\!f}}{R^*}\right) \bm{\hat e_x}
    = \frac{1}{2} \rho V_{LSC}  U_{f\!f} R^{*}\ \bm{\hat e_x}.
    \label{equ:angularmomentum}
\end{equation}

\subsection{Thermoelectric Currents at the Electrically Conducting Boundaries}
\begin{figure}
  \centering
    \includegraphics[width=0.9\textwidth]{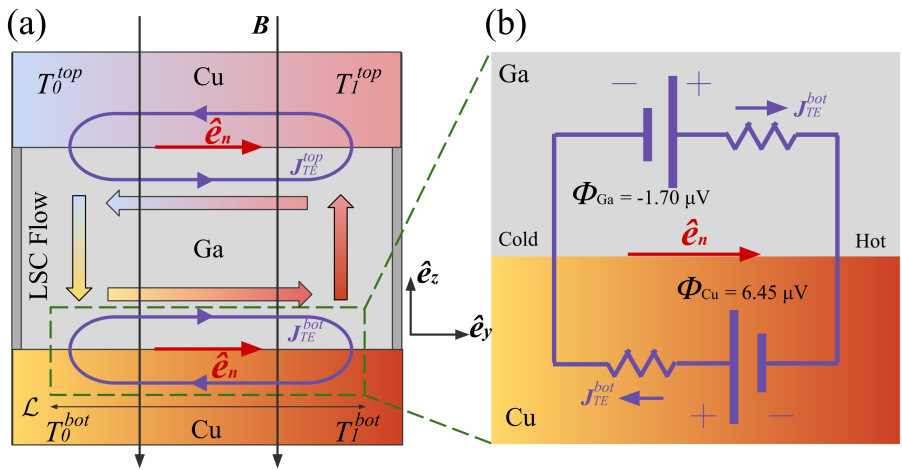}
    \caption{(a) Cross-sectional schematics of the experimental MC system with electrically conducting boundaries. 
In the plane of LSC, the turbulent LSC imprints large scale thermal anomalies onto the boundaries: the top boundary has a minimum temperature $T_{0}^{top}$ and a maximum temperature $T_{1}^{top}$; the bottom boundary has a minimum temperature $T_{0}^{bot}$ and a maximum temperature $T_{1}^{bot}$. Thermoelectric \blue{potentials} are generated at \JA{the Cu-Ga interfaces and form current density loops across the boundaries, $\boldsymbol J_{T\!E}^{top}$ and $\boldsymbol J_{T\!E}^{bot}$, with a width of $\mathcal L \approx \Gamma H$. (b) Circuit diagram of the Cu-Ga system at the bottom boundary. The thermoelectric \blue{potential} in gallium is denoted as $\Phi_{Ga}$, which is smaller in magnitude and has an opposite sign as the thermoelectic \blue{potential} in copper, $\Phi_{Cu}$.} Thus, the thermoelectric current flows from cold to hot in liquid gallium (in $+\bm{\hat e_n}$), and from hot to cold in copper (in $-\bm{\hat e_n}$). }
    \label{fig:circuit}
\end{figure}

Figure \ref{fig:circuit}(a) shows a schematicized vertical slice through our experimental tank in the low $N_{\mathcal{C}}$ regime. The LSC generates horizontal thermal gradients on both horizontal boundaries. Thus, the end-blocks have a higher temperature near the upwelling branch of the LSC, which carries warmer fluid upwards, and the end-blocks are cooler near the downwelling branch of the LSC, which carries cooler fluid downwards. These temperature gradients on the top and bottom fluid-solid interfaces generate thermoelectric current loops. \JA{Eq.} (\ref{equ:TEemfb}) is used to calculate the net Seebeck coefficient of such a thermoelectric current loop in our Cu-Ga system:
\begin{equation}   
    \Phi_{T\!E} = \int_{r_0(T_{0})}^{r_1(T_{1})} (S_{Cu}-S_{Ga}) \boldsymbol \nabla T \cdot d \boldsymbol r = \frac{\pi^{2} k_B^{2}}{6 e}\left[\frac{x_{Cu}}{E_{F Cu}}-\frac{x_{Ga}}{E_{F Ga}}\right] \left(T_1^{2}-T_{0}^{2}\right),
    \label{equ:Phi_TE}
\end{equation}   
where $S_{Ga}$ and $S_{Cu}$ are Seebeck coefficients for gallium and copper, respectively, calculated via eq. (\ref{equ:seebeck}). For gallium, the numeric coefficient $x_0$ is $x_{Ga} = 0.7$ \citep{cusack1963electronic} and the Fermi energy is $E_{FGa} = 10.37\ \mathrm{eV}$ \citep{kasap2001thermoelectric}. For copper, $x_{Cu} = -1.79$ and $E_{FCu} = 7.01\ \mathrm{eV}$. The temperatures $T_0$ and $T_1$ represent the minimum and maximum temperatures, respectively, in a given horizontal plane (e.g., figure \ref{fig:TE_AB}). 

Following eq. (\ref{equ:SeebeckAB}), the net Seebeck coefficient on the $k$-level Cu-Ga interface is:
\begin{equation}      
    \widetilde{S}^k = \frac{\pi^{2} k_B^{2}}{3 e}\left[\frac{x_{C u}}{E_{F C u}}-\frac{x_{G a}}{E_{F G a}}\right] T^k \equiv X_0 \, T^k \ [\mathrm{V/K}],
    \label{equ:S_Cu-Ga}
\end{equation}   
where $T^k $ is the time-azimuthal mean temperature on the $k$-interface. The Cu-Ga Seebeck prefactor $X_0$ collects all the constant and material properties in eq. (\ref{equ:S_Cu-Ga}). Its value in our system is
\begin{equation} 
X_0 \approx -7.89 \times 10^{-9}\ \mathrm{V/K^2}. 
\end{equation}
Unlike the net Seebeck coefficient, $X_0$ does not depend on temperature.

The thermoelectric current density vector in liquid gallium is approximated via eq. (\ref{equ:J_TE}):
\begin{equation}
    \boldsymbol J_{T\!E}^k  \approx \sigma_0 X_0 T^k  \left( \frac{\delta T^k}{\mathcal L} \right)  \, \bm{\hat e_y}, 
    \label{equ:seebeck_current2}
\end{equation}
where $\sigma_0 = 3.63\times 10^6\ \mathrm{S/m}$ is the effective electric conductivity for the Cu-Ga system, calculated by substituting \JA{the} conductivity of gallium ($\sigma_{Ga}\approx 3.88\times 10^6\ \mathrm{S/m}$) and copper ($\sigma_{Cu}\approx 5.94\times 10^7\ \mathrm{S/m}$) into eq. (\ref{equ:effelec}). The horizontal temperature gradient is approximated by the maximum temperature difference across the $k$-interface, $\delta T^k$, divided by a characteristic width of the current loop, $\mathcal L $. We assume this width is the same as the diameter \JA{$D$} of the tank, \JA{$\mathcal L \approx \Gamma H = 2H = 197.2\ \mathrm{mm}$. (Effects of possible TE currents in the stainless steel sidewall ($\sigma_{St.Stl.}\approx 1.4\times 10^6\ \mathrm{S/m}$) are not accounted for here.)}

\blue{Figure \ref{fig:circuit}(b) shows a circuit diagram for the thermoelectric current loop near the \JA{experiment's} bottom liquid-solid interface at \JA{$\approx 40^{\circ}\, \mathrm C$.} The thermoelectric potential in gallium has a negative sign, \blue{so} the currents within the fluid are always aligned in the direction of the thermal gradient $\bm{\hat e_n} = \bm{\hat e_y}$. In contrast, the thermoelectric potential in copper has a positive sign, so the current flows from the hot to the cold region in $-\bm{\hat e_n}$.} 

\subsection{Thermoelectric Forces and Torques}
\begin{figure}
  \centering
    \includegraphics[width = 0.9\textwidth]{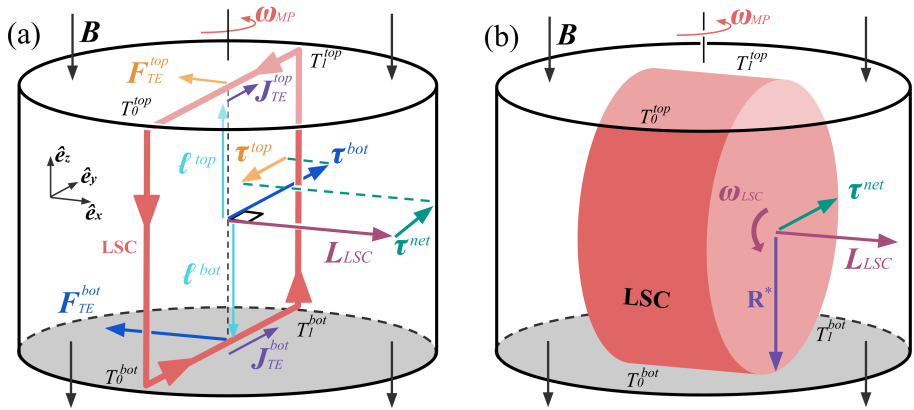}
    \caption{(a) \JA{Free-body diagram of thermoelectric LSC precession. The red arrows enclosed inside the tank represents the LSC. The thermoelectric potentials generate current in the liquid gallium at top and bottom boundaries: $\boldsymbol J_{T\!E}^{top}$ and $\boldsymbol J_{T\!E}^{bot}$ with $J_{T\!E}^{top}<J_{T\!E}^{bot}$. The thermoelectric Lorentz forces, $\boldsymbol  F_{T\!E}^{top}$ and $\boldsymbol F_{T\!E}^{bot}$, create a net torque $\bm \tau_{net}$ perpendicular to the LSC's angular momentum vector, $\bm L_{LSC}$, which drives the LSC to precess around the tank's vertical $\bm{\hat e_z}$-axis. (b) Precessional flywheel schematic (not to scale). The LSC is simplified into a flywheel-like cylinder of radius $R^*$, with angular velocity $\omega_{LSC} \lesssim U_{f\!f}/R^*$ and angular momentum $\bm L_{LSC} = \omega_{LSC} MR^{*2} \bm{\hat e_x}/ 2$. The LSC is assumed to respond to $\bm \tau_{net}$ in a solid-body manner. }}
    \label{fig:flywheel}
\end{figure}

The thermoelectric component of the mean Lorentz forces density on the index $k$ horizontal interface can be calculated, using  eq. (\ref{equ:seebeck_current2}),  as
\begin{equation}
    \boldsymbol f_{T\!E}^k  
    =  \boldsymbol J_{T\!E}^k \times \boldsymbol B =  \sigma_0 X_0 T^k \left(\frac{\delta T^k }{\JA{\Gamma}H} \bm{\hat e_y}\right) \times B  \bm{\hat e_b} 
       = \frac{\sigma_0 X_0 B \,  T^k  \delta T^k   }{\JA{\Gamma} H} \, (\bm{\hat e_y} \times \bm{\hat e_b}) \, ,  
    \label{equ:lorentzdensity}
\end{equation}
where we have taken $\mathcal{L} \approx \Gamma H$. Since the thermoelectric currents are predominantly in the \JA{LSC plane} and are aligned parallel to the thermal gradient, the thermoelectric Lorentz forces point in the $+\bm{\hat e_x}$-direction for upward directed $\bm{B}$, and in the $-\bm{\hat e_x}$-direction for downward directed $\bm{B}$ (corresponding to figure \ref{fig:flywheel}(a)). 

The Lorentz force is the volume integral of the force density:
\begin{equation}
    \bm F_{T\!E}^k = \int{\bm f_{T\!E}^k \ \mathrm{dV}} = \int \boldsymbol J_{T\!E}^k \times \boldsymbol{B}\ \mathrm{dV}.
\end{equation}
%
We coarsely assume that each thermoelectric current loop exists in the top half or bottom half of the LSC, and generates uniform Lorentz forces that act, \JA{respectively,} on the upper or lower half of the LSC. With these assumptions, we take each hemicylindrical integration volume to be $V_{LSC}/2$. The Lorentz forces due to thermoelectric currents generated across the $k$-level Cu-Ga interface are then estimated to be
\begin{equation}
      \boldsymbol F_{T\!E}^k =  \frac{1}{2} V_{LSC} \bm f_{T\!E}^k  \, . 
     \label{equ:lorentz}
\end{equation}

In order to estimate the torques due to each thermoelectric current loop, we assume that the thermoelectric Lorentz forces act on the LSC via a moment arm $\bm \ell$ of approximate length $H/2$.  Thus, $\bm \ell^{top} = H/2 \, \bm{\hat{e}_z}$ and $\bm \ell^{bot} = - H/2 \, \bm{\hat{e}_z}$. 
The net thermoelectric torque on the LSC then becomes
\begin{subequations}
\begin{eqnarray}
    \bm \tau^{net}  &=& \bm \tau^{top}+ \bm \tau^{bot} \\
    &=& \left( \bm \ell^{top} \times  \boldsymbol F_{T\!E}^{top} \right) + \left( \bm \ell^{bot} \times  \boldsymbol F_{T\!E}^{bot} \right) \\
    &=& (H/2) \bm{\hat e_z} \times \left( \boldsymbol F_{T\!E}^{top}- \boldsymbol F_{T\!E}^{bot} \right) \\
    &=& (H V_{LSC}/4) \bm{\hat e_z} \times \left( \bm f_{T\!E}^{top}- \bm f_{T\!E}^{bot} \right)\, .     
    \label{Eq:nettorque.d}
\end{eqnarray}
\end{subequations}
Substituting eq. (\ref{equ:lorentzdensity}) into eq. (\ref{Eq:nettorque.d}) yields the net thermoelectric torque on the LSC to be
\JA{
\begin{eqnarray}
\label{equ:nettorque2}
\bm \tau^{net} &=& \frac{\sigma_0  V_{LSC}  X_0 B \, \big( T^{bot} \delta T^{bot} -  T^{top} \delta T^{top} \big)}{4 \Gamma}  \    \big(\bm{\hat e_x} \times \bm{\hat e_b} \big)  \\
&=& \frac{\sigma_0  V_{LSC}  X_0 B \, \mathcal{T}}{4 \Gamma}  \,   \big(\bm{\hat e_x} \times \bm{\hat e_b} \big) \, ,\nonumber
\end{eqnarray}
}where 
\begin{equation}
\mathcal{T} \equiv \big( T^{bot} \delta T^{bot} -  T^{top} \delta T^{top} \big)
\label{equ:mathcalT}
\end{equation}
describes the difference in thermal conditions on the bottom relative to top horizontal Cu-Ga interfaces.
Since $T^{bot} > T^{top}$ in all our convection experiments, the data in figure \ref{fig:delta_deltaT}(b) implies that $\mathcal{T} > 0$ in the MP regime. 
Since all the other parameters in eq. (\ref{equ:nettorque2}) are positive, the net thermoelectric torque is directed in $-\bm{\hat e_y}$ for upwards directed magnetic fields ($\bm{\hat e_b} = + \bm{\hat e_z}$) and, as shown in figure \ref{fig:flywheel}, the net torque is directed in $+\bm{\hat e_y}$ for downwards directed magnetic fields ($\bm{\hat e_b} = - \bm{\hat e_z}$).  \JA{Eq. (\ref{equ:nettorque2}) also shows that the bottom torque will tend to dominate even when $\delta T^{bot} \approx \delta T^{top}$ since $T^{bot} > T^{top}$ in all convectively unstable cases.}

\begin{table}
    \centering
    \def~{\hphantom{0}}
    \begin{tabular}{lll}
\textbf{Symbols} & \textbf{Description} & \textbf{Value} \\[5pt]
$\sigma_0$         & Cu-Ga effective electric conductivity, eq. (\ref{equ:effelec})    & $3.63\times10^6\ \mathrm{S/m}$    \\[3pt]
$B$            &  magnetic field intensity                & $120\ \mathrm{\blue{gauss}}$                \\[3pt] 
$\mathcal L$    & Horizontal length scale of TE current loops, \JA{$\approx \!\Gamma H$}    & $197.2\ \mathrm{mm}$        \\[3pt]
$X_0$            & Cu-Ga Seebeck prefactor, eq. (\ref{equ:S_Cu-Ga})           & $-7.89 \times 10^{-9}\ \mathrm{V/K^2}$ \\ [3pt]
$\rho$           & Liquid gallium density                & $6.08 \times 10^3\ \mathrm{kg/m^3}$ \\ [3pt]
$R^*$            & Effective radius of the LSC, eq. (\ref{equ:Rstar})             & $0.08\ \mathrm m$                 \\ [3pt]
$U_{f\!f}$       & Free-fall velocity, eq. (\ref{Eq:ff})                    & $0.03\ \mathrm{m/sec}$      \\   [3pt]
$\overline{T}$     & Mean fluid temperature, eq. (\ref{Eq:MeanT})  & $42.50\ \mathrm {^\circ C}$  \\[3pt]
$\Delta T$     & Vertical temperature difference across the fluid, eq. (\ref{Eq:DeltaT})  & $7.03\ \mathrm K$  \\[3pt]
$T^{bot}$       & Bottom interface mean temperature   & $319.23\ \mathrm K$               \\[3pt]
$\delta T^{bot}$     & Bottom interface mean temperature difference, eq. (\ref{equ:deltaT_h2})  & $3.44\ \mathrm K$  \\[3pt]
$T^{top}$       & Top interface mean temperature      & $312.07\ \mathrm K$               \\ [3pt]
$\delta T^{top}$     & Top interface mean temperature difference, eq. (\ref{equ:deltaT_h2})     &$2.24\ \mathrm K$  \\ [3pt]
$\mathcal{T}$       & $(T^{bot}\delta T^{bot} - \delta T^{top}\delta T^{top})$      & $399.11 \ \mathrm K^2$    \\ [3pt]
    \end{tabular}
    \caption{Experimental parameter values from the $Conducting\ MC^-$ subcase. These values are characteristic of those used in \JA{calculating} $\omega_{M\!P}$ in figure \ref{fig:omega}.}
    \label{tablePMpara}
\end{table}

\subsection{Thermoelectrically-driven LSC Precession}

The net thermoelectric torque on the LSC acts in the direction perpendicular to $\bm{L}_{LSC}$. The LSC must then undergo a precessional motion in order to conserve angular momentum.  This precession can be quantitied via Euler's equation \citep{landau1976theoretical}, in which the net torque is the time derivative of the angular momentum: 
\begin{equation}
    \boldsymbol \tau^{net} = \frac{d{\bm L}_{LSC}}{d t} = I\  \frac{d {\boldsymbol{\omega}}}{d t}+\boldsymbol{\omega} \times \bm L_{LSC},
    \label{equ:Euler}
\end{equation}
The angular velocity vector $\boldsymbol{\omega}$ is comprised of two components here 
\begin{equation} 
\boldsymbol{\omega} = \omega_{LSC} \bm{\hat e_x} + \boldsymbol \omega_{M\!P} \, ,
\end{equation}
where $\omega_{LSC}$ is the angular velocity component of the flywheel in $\bm{\hat e_x}$ and $\boldsymbol \omega_{M\!P}$ is the angular velocity vector of the LSC's magnetoprecession. We assume that the precession frequency and the angular speed of the flywheel are nearly time-invariant, $d {\boldsymbol{\omega}}/{d t} \approx 0$. Then eq. (\ref{equ:Euler}) reduces to
\begin{equation}
    \bm{\tau}^{net} = \left( \omega_{LSC} \bm{\hat e_x} + \boldsymbol \omega_{M\!P}  \right) \times  L_{LSC} \bm{\hat e_x} =  \bm{\omega}_{M\!P}  \times  L_{LSC} \bm{\hat e_x}\, ,
    \label{equ:Euler2}
\end{equation}
where $\bm \omega_{LSC} \times \bm L_{LSC} = 0$ since these vectors are parallel. Expression eq. (\ref{equ:Euler2}) requires that $\boldsymbol \omega_{M\!P}$ remain orthogonal to both $\boldsymbol L_{LSC}$ and $\boldsymbol \tau_{net}$ such that 
\begin{equation}
    {\tau}^{net} \big(\bm{\hat e_x} \times \bm{\hat e_b} \big) =   \bm{\omega}_{M\!P}  \times  L_{LSC} \bm{\hat e_x} 
    =    {\omega}_{M\!P}  L_{LSC}  \, \big( \bm{\hat{\omega}}_{M\!P}  \times  \bm{\hat e_x} \big) \, .
    \label{equ:Euler3}
\end{equation}
Therefore, $\bm{\hat{\omega}}_{M\!P} = - \bm{\hat e_b}$ and the precessional angular velocity vector is 
\begin{equation}
    \bm \omega_{M\!P} = \frac{ \tau_{net}}{L_{LSC}} \big( \! - \bm{\hat e_b} \big) \, .
    \label{equ:angular}
\end{equation}
Substituting eqs. (\ref{equ:angularmomentum}) and (\ref{equ:nettorque2}) into eq. (\ref{equ:angular}) yields our analytical estimate for the thermoelectrically-driven angular velocity of LSC precession in the $N_{\mathcal{C}} \lesssim 1$ TEMC experiments: 
\blue{
\begin{subequations}
 \label{equ:pmtheory}
    \begin{eqnarray}
    \boldsymbol \omega_{M\!P}
    &=& \frac{\sigma_0 X_0 B}{2 \rho  U_{f\!f}\Gamma R^{*} } \big( T^{bot} \delta T^{bot} -  T^{top} \delta T^{top} \big) \big( \! - \bm{\hat e_b} \big) \\
     &=&    \frac{\sigma_0 X_0 B \mathcal{T} }{2\rho U_{f\!f} \Gamma R^{*} } \big( \! - \bm{\hat e_b} \big) \\
     &=&    \frac{\pi^{1/2}}{\Gamma^{3/2}} \frac{\sigma_0 X_0 B \mathcal{T} }{2  \rho U_{f\!f} H } \big( \! - \bm{\hat e_b} \big)\, . 
    \end{eqnarray}
\end{subequations}
}
Expressions (\ref{equ:angular}) and eq. (\ref{equ:pmtheory}) predict that the LSC's magnetoprecessional angular velocity vector, $\bm \omega_{M\!P}$, will always be antiparallel to the imposed magnetic field vector in our Cu-Ga TEMC experiments in which $\mathcal{T} > 0$. Further, the magnetoprecession should flip direction such that $\bm \omega_{M\!P}$ would be parallel to $\bm{B}$ in a comparable Cu-Ga TEMC system with $\mathcal{T} < 0$. These predictions agree with the precession directions found in our MP regime experiments, as shown in figure \ref{fig:bflip}(a). 

\subsection{Experimental Verification}
\label{sec:exp_veri}

The direction of magnetoprecession is sensitive to the sign of $\mathcal{T}$. The value and sign of $\mathcal{T}$ are, however, both likely related to the details of the experimental set up. For instance, we have a thermostated bath controlling the top thermal block temperature, whereas a fixed thermal flux is input below the bottom thermal block.  Further, the top and bottom thermal blocks have different thicknesses. It is possible that $\mathcal{T}$ could behave differently with much thinner or thicker end-blocks, and for differing thermal boundary conditions. Thus, further modeling efforts are required before $\mathcal{T}$ can be predicted \textit{a priori}. 

\begin{figure}
  \centering
    \includegraphics[width = \textwidth] {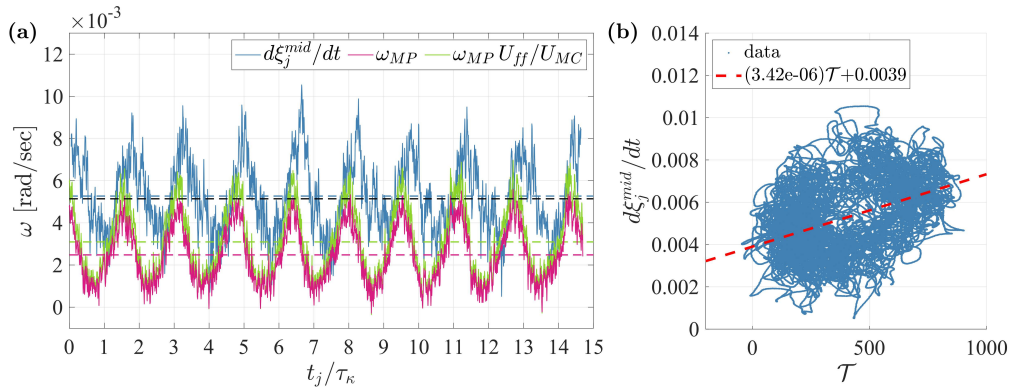}
    \caption{(a) Time series of LSC precessional rate for the $Conducting\ MC^-$ case at $Ra = 1.83\times 10^6$, $Ch = 2.59\times10^3$, and $N_\mathcal{C}= 0.31$. The horizontal axis shows time $t$ normalized by the diffusion timescale $\tau_\kappa$. The vertical axis is the LSC's instantaneous angular precession speed $\omega$. The \blue{blue} line shows angular velocity of the LSC plane, $d \xi^{mid}_j / dt$, measured via eq. (\ref{equ:Tfit}) using temperature data on the midplane sidewall. The magenta line marks $\omega_{M\!P}$ model predictions made using eq. (\ref{equ:pmtheory}) and instantaneous temperature data. The green line shows the alternative MP prediction made using eq. (\ref{equ:pmtheory}) with \citet{zurner2020flow}'s $U_{M\!C}$ velocity scaling eq. (\ref{equ:Zurner}). See Appendix Table \ref{app:table1} for $Conducting\ MC^-$ case details. The horizontal dashed lines are mean values for their corresponding angular speeds. The black horizontal dashed line represents the peak frequency from the FFT converted into the angular speed. (b) Scatter plot of $d \xi^{mid}_j / dt$ versus $\mathcal T$. 
The red dashed line is a linear fit for this particular case. The best fit slope, $3.42\times 10^{-6}\, \mathrm{K^2/s}$, is $55.3\%$ of the theoretical prediction from (\ref{equ:pmtheory}), where the prefactor $\sigma_0 X_0 B / ( 4\rho U_{f\!f} R^{*} ) \approx 6.18\times 10^{-6}\, \mathrm{K^2/s}$.}
    \label{fig:omega}
\end{figure}

Figure \ref{fig:omega}\blue{(a)} provides a test of our magnetoprecession model using data from the $Conducting\ MC^-$ case. The blue line shows the experimentally determined LSC angular velocity via measurements of the azimuthal drift velocity of the LSC plane, $d \xi^{mid}_j / dt$, made with the best fits to eq. (\ref{equ:Tfit}). The magenta line shows the instantaneous angular velocity of LSC precession, $(\omega_{M\!P})_j$, calculated by feeding instantaneous thermal data from the $Conducting\ MC^-$ case into eq. (\ref{equ:pmtheory}). The green line is a modified theoretical estimate, in which the $U_{M\!C}$ velocity prediction (\ref{equ:Zurner}) is used in eq. (\ref{equ:pmtheory}) instead of the upper bounding $U_{f\!f}$ estimate.
 
The three time series in figure \ref{fig:omega}\blue{(a)} show that the predicted magnetoprecessional angular speeds compare well with the LSC azimuthal drift speed, especially when accounting for the simplifications that are underlying eq. (\ref{equ:pmtheory}). The time series all have similar gross shapes, but with the peaks in $d \xi^{mid}_j /dt$ slightly delayed in time relative to those in the $\omega_{M\!P}$ curves. The dashed lines in figure \ref{fig:omega} show the time-mean angular velocity values. The time mean LSC plane velocity is $d \xi^{mid} /dt = 5.27\times 10^{-3}\ \mathrm{rad/s}$.  The mean value of the magnetoprecessional flywheel (magenta) is $\omega_{M\!P} = 2.48\times 10^{-3}\ \mathrm{rad/s} = 0.47 \, d \xi^{mid} /dt$.  The mean value of the modified estimate (green) is $\omega_{M\!P}\, U_{f\!f}/ U_{M\!C}= 3.10\times 10^{-3}\ \mathrm{rad/s} = 0.59 \, d \xi^{mid} /dt$. Moreover, there is a good agreement between the converted angular speed from the peak frequency $2\pi f_{peak}$ of \JA{the} FFTs (black dashed lines) and the mean LSC azimuthal drift speed by direct thermal measurement $d \xi^{mid} /dt$ (blue dashed lines). The direct measurement is about $2.5\%$ higher than $2\pi f_{peak}$.  

\blue{Figure \ref{fig:omega}(b) shows the correlation between the measured angular velocity $d \xi^{mid}_j /dt$ and $\mathcal{T}$. The angular velocity values correspond to the blue curve in panel (a). The dashed red line is the best linear fit to the data, and shows that there is a net positive correlation between $d \xi^{mid}_j /dt$ and the asymmetry of the thermal condition in end-blocks, represented by $\mathcal T$. The best fit slope is $(d \xi^{mid}_j /dt) / \mathcal T = 3.42\times 10^{-6}\, \mathrm{K^2/s}$. Theoretical prediction (\ref{equ:pmtheory}) gives $\omega_{M\!P}/ \mathcal T = \sigma_0 X_0 B / ( 4\rho U_{f\!f} R^{*} ) = 6.18\times 10^{-6}\, \mathrm{K^2/s}$. Thus, the best-fit slope agrees with theory to within approximately a factor of two. }

\begin{figure}
  \centering
    \includegraphics[width=0.8\textwidth]{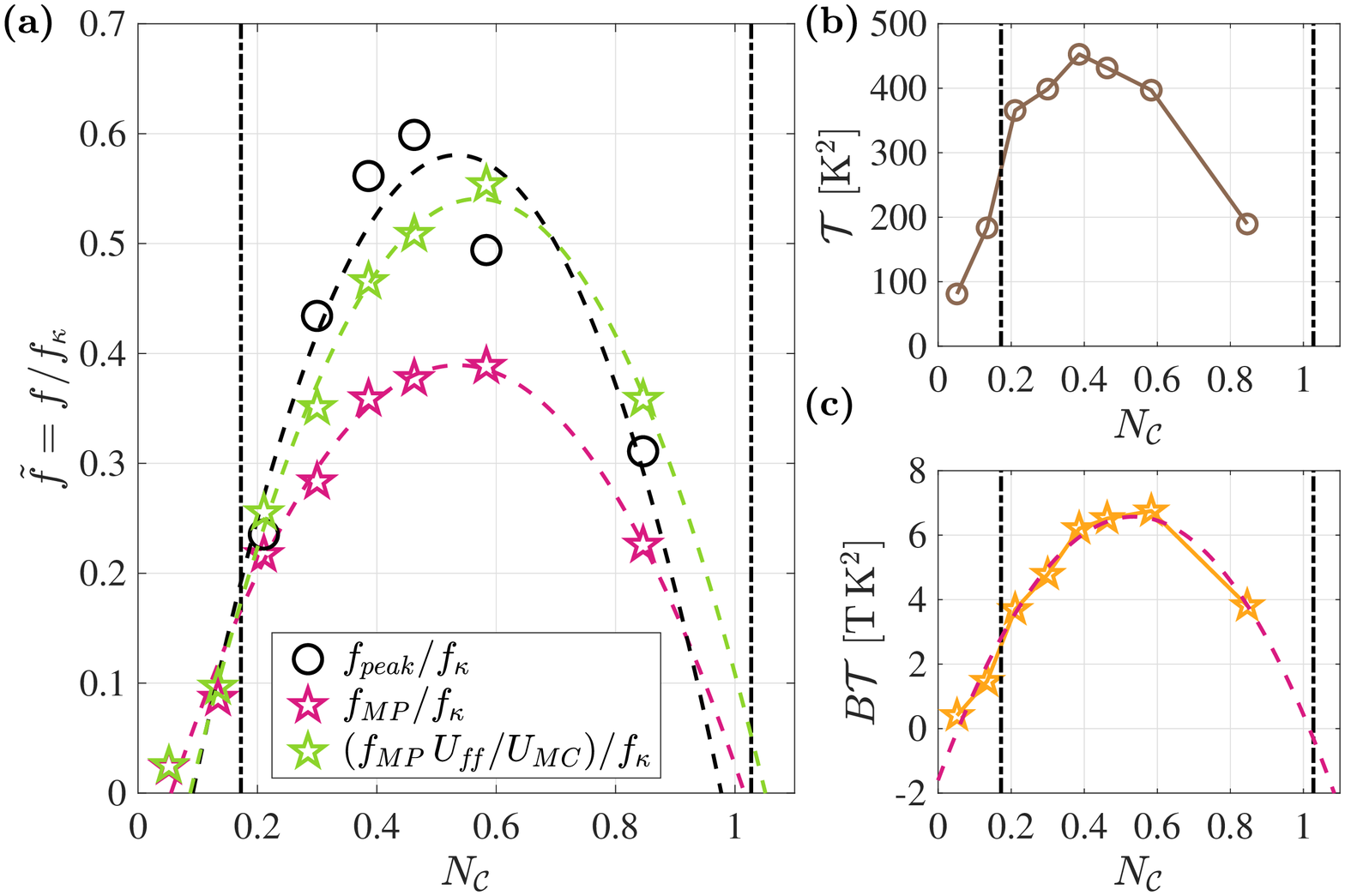}
    \caption{ (a) Normalized precessional frequency, $\tilde f = f/f_\kappa$, versus convective interaction parameter, $N_{\mathcal{C}}$, where the thermal diffusion frequency is $f_\kappa \equiv 1/\tau_\kappa \approx 1.34\times 10^{-3}$ Hz. Black circles denote the peak frequencies of FFT spectra for the fixed $Ra \approx 2 \times 10^6$ experimental survey. Magenta stars are frequencies predicted by the TEMC precession model, $f_{M\!P}/f_\kappa = \omega_{M\!P}/(2 \pi f_\kappa)$. Green stars correspond to $\omega_{M\!P} U_{f\!f} / (2 \pi U_{M\!C}f_\kappa)$, the frequency of the precession model using the $U_{M\!C}$ scaling velocity. Dashed curves represent the second-order best fit curves. (b) $\mathcal{T}$ versus $N_{\mathcal{C}}$. 
    (c) The product $B \mathcal{T}$ versus $N_{\mathcal{C}}$. The magneta dashed curve is the second degree best fit curve from panel (a) normalized by the factor $\sigma_0 X_0/ (8\pi f_\kappa \rho U_{f\!f} R^*) \approx 0.0612\, [\mathrm{T^{-1}K^{-2}s^{-1}}]$.}  
    \label{fig:theorydatafit}
\end{figure}

The thermoelectric LSC precession model is tested further  in figure \ref{fig:spg}(a) and figure \ref{fig:theorydatafit} using the fixed $Ra \approx 2 \times 10^6$ case results. In figure \ref{fig:spg}(a), the peak of each experiment's thermal FFT spectrum is marked by an \blue{open black} circle.  The circles in the MP \blue{regime} are connected by a best fit parabola (black dashed curve). The white stars show the predicted magnetoprecessional frequency, $f_{M\!P} = \omega_{M\!P}/2 \pi$, calculated using eq. (\ref{equ:pmtheory}). The white dotted line is the best parabolic fit to the $f_{M\!P}$ values. Adequate agreement is found between the spectral peaks and the $f_{M\!P}$ values. The low frequency tail of the best fits appears to correlate with secondary peaks in the thermal spectra.  This suggests that MP modes exist down to $Ch \simeq 10^3$ in the JRV \blue{regime}, but with weaker spectral signatures that do not dominate those of the jump rope LSC flow.

Figure \ref{fig:theorydatafit}(a) shows a linear plot of FFT spectral peaks (black circles), $f_{M\!P}$ predictions (magenta stars), and the modified predictions $f_{M\!P} (U_{f\!f}/U_{M\!C})$ (green stars) plotted versus $N_{\mathcal{C}}$ for the fixed $Ra \approx 2 \times 10^6$ MC experiments with $N_\mathcal{C} < 1$.  The dashed lines show best parabolic fits.  The vertical dot-dashed lines are the regime boundaries that separate the JRV, MP and MCMC \blue{regimes} in figure \ref{fig:spg}(a).  The $f_{M\!P}$ estimates differs by $\lesssim 37\%$ from the experimental spectral data, while the improved model that makes use of $U_{M\!C}$ differs from the spectral peak frequencies by $\lesssim 16\%$.  Further, the intersections of the best fit parabolas with $f/f_\kappa = 0$, near $N_\mathcal{C} \approx 0.1$ and $N_\mathcal{C}\approx 1$,  agree well with the empirically located regime boundaries.  

Figure \ref{fig:theorydatafit}(b) shows $\mathcal{T} =  ( T^{bot} \delta T^{bot} -  T^{top} \delta T^{top} )$ versus $N_{\mathcal{C}}$ for the experimental cases shown in panel (a).  Although it resembles the curves in panel (a), the shape of the $\mathcal{T}$ curve is steeper on its $N_{\mathcal{C}} < 0.4$ branch and its peak is shifted to slightly lower $N_{\mathcal{C}}$. Figure \ref{fig:theorydatafit}(c) shows the product $B \mathcal{T}$ plotted in orange as a function of  $N_{\mathcal{C}}$. The magenta dashed curve in Figure \ref{fig:theorydatafit}(c) is the best fit parabolic curve from panel (a) normalized by $8\rho U_{f\!f}R^*/(\sigma_0 X_0)$, which, according to eq. (\ref{equ:pmtheory}), separates these values.  By comparing figures \ref{fig:theorydatafit}(b) and \ref{fig:theorydatafit}(c), we argue that the quasi-parabolic structure of the MP frequency data in the MP \blue{regime} is controlled by trade-offs between the $B$ and $\mathcal{T}$ trends.

%% file: sections/6_Discussion.tex
We have conducted a series of magnetoconvection (MC) laboratory measurements in turbulent liquid gallium convection with a vertical magnetic field and thermoelectric (TE) currents in cases with electrically-conducting boundaries. Three regimes of TEMC flow are found: i) the LSC sustains its flow structure in the $0< \Nc \lesssim 0.1$ jump rope vortex (JRV) regime; ii) long-period magnetoprecession (MP) of the LSC dominates the $0.1\lesssim \Nc\lesssim 1$ MP regime; and iii) the LSC is replaced by a multi-cellular magnetoconvective flow pattern in the $\Nc \gtrsim 1$ MCMC regime.
 
\begin{figure}
  \centering
    \includegraphics[width =0.8\textwidth]{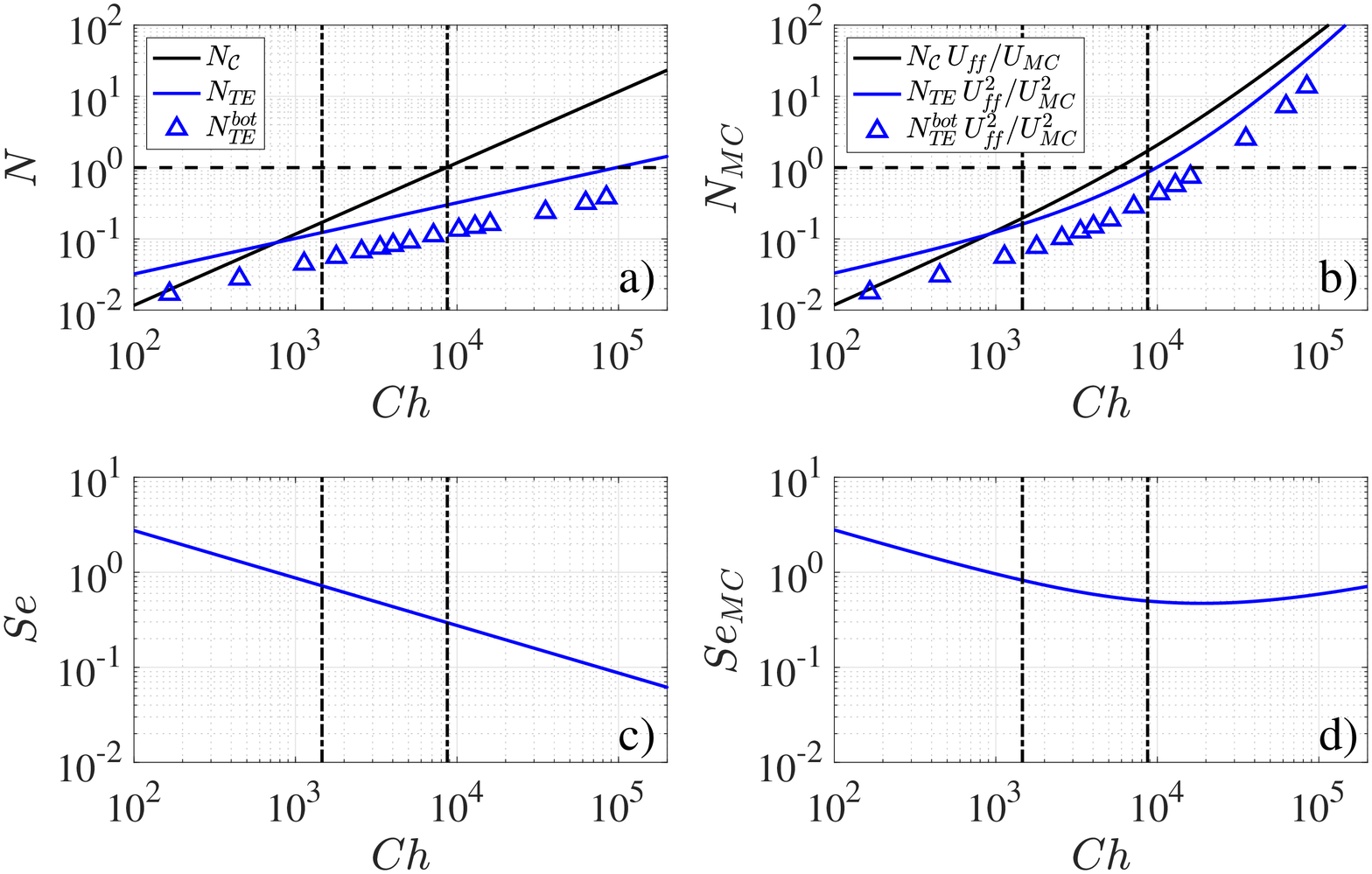}
    \caption{ a) Convective and thermoelectric interaction parameters, $\Nc$ and $\Nte$, plotted versus Chandrasekhar number $Ch$ \blue{over the parameter space of the fixed-$Ra$ survey}. The black line shows $\Nc = \sigma B^2 H/(\rho U_{f\!f})$, and the blue line shows $\Nte = \sigma B |\widetilde S| \Delta T/(\rho U_{f\!f}^2)$, \JA{following the definitions} in eq. (\ref{equ:defNc}) and eq. (\ref{equ:defNTE}). The net Seebeck coefficient $ \widetilde S = X_0 \overline T$ is defined using the mean temperature of the system. The blue triangles denote $\Nte^{bot}= \sigma B |\widetilde S| \delta T^{bot}/(\rho U_{f\!f}^2)$ \blue{calculated for the experimental data of the fixed-$Ra$ cases. The two vertical black dot-dashed lines separate the parameter space into the JRV, MP and MCMC regimes from left to right in each panel.} Panel (b) is comparable to (a), but employs the characteristic MC flow speed $U_{MC}$ in place of $U_{f\!f}$. Panels (c) and (d) show the corresponding Seebeck numbers, $Se = \Nte/\Nc$ and $Se_{MC} = (\Nte/\Nc) \, (U_{f\!f}/U_{MC})$, respectively.
}
    \label{fig:NTE_NC}
\end{figure}
Figure \ref{fig:NTE_NC}(a) shows the convective and thermoelectric interaction parameters, $\Nc$ and $\Nte$ respectively, as a functions of $Ch$. The vertical dashed lines separate the parameter space into the three characteristic regimes in figure \ref{fig:spg}, with JRV on the left hand side, MP in the middle and MCMC on the right hand side. Both $\Nc$ and $\Nte$ are approximately of the same order in the MP regime. 
\blue{The blue open triangles \JA{correspond to} the experimentally derived thermoelectric interaction parameters at the bottom boundary for the fixed-$Ra$ survey, $\Nte^{bot}= \sigma B |\widetilde S| \delta T^{bot}/(\rho U_{f\!f}^2)$. The bottom boundary temperature data, $\delta T^{bot}$, are used here because it has the larger $\Nte$ than the top layer, and dominates the dynamics of the magnetoprecession. }

\blue{Figure \ref{fig:NTE_NC}(b)} uses $U_{MC}$ as the characteristic flow speed instead of $U_{f\!f}$. Similar to (a), both $\Nc$ and $\Nte$ are of the same order and roughly aligned with each other in the MP regime, which means the quasistatic Lorentz forces from the fluid motions are comparable to the thermoelectric Lorentz forces. Panels (c) and (d) show $Se$ and $Se_{MC}$, which are the ratios of the blue and black lines in panels (a) and (b), respectively. Since the convective and TE interaction parameters are order unity in the MP regime in figure \ref{fig:NTE_NC}, this shows that both the Lorentz forces are approximately comparable to the buoyant inertia. Thus, a triple balance is possible between the motionally-induced Lorentz forces, the thermoelectric Lorentz forces and the buoyancy forces in the MP regime. 

Our thermoelectrically-driven precessional flywheel model provides an adequate characterization of the MP mode that is observed in turbulent MC experiments with electrically conducting boundaries. The model predicts the zeroth-order precessional frequencies of the MP mode.  Further, it explains the changing direction of precession when the imposed magnetic field direction is reversed (figure \ref{fig:bflip}). 

\JA{There are, however,} a number of limitations to our flywheel model. First, we have allowed Ga to corrode the \blue{Cu} end-blocks in order to ensure good material contact across the interfaces.  The reaction between \blue{Cu} and \blue{Ga} forms a gallium alloy layer on the copper boundary.  This ongoing reaction should decrease the interfacial electrical conductivity over time, resulting in a smaller net torque on the LSC and a slower rate of magnetoprecession. This may explain a subtle feature in figure \ref{fig:spg}: the peak frequency from the $MC^-$ case is higher than the comparable fixed-$Ra$ survey case that was carried out months later but with similar control parameters. Thus, Ga-Cu chemistry at the interface appears to matter in the TE dynamics, yet we are currently unable to \JA{control or to parameterize these interfacial reaction processes.} Second, our model requires measurements of the horizontal temperature gradients in the conducting end-blocks.  A fully self-consistent model would use the input parameters to predict these gradients {\it a priori}, independent of the experimental data.

A long-period precessional drift of the LSC has also been observed in water-based laboratory experiments influenced by the Earth's rotation \citep{brown2006effect}. In \JA{their} experiments, the LSC rotates azimuthally with a period of days. This period is over two orders of magnitude greater than our MP period. \JA{Since MP modes are found only to develop in the presence of imposed vertical magnetic fields and electrically-conducting boundaries, it is not possible to explain the MP mode solely due to Coriolis effects from Earth's rotation.}  

Alternatively, the rotation of the Earth could couple with the magnetic field to induce magneto-Coriolis (MC) waves in the convecting gallium \citep[e.g.,][]{finlay2008course, hori2015slow, schmitt2013magneto}. In the limit of strong rotation, MC waves have a slow branch that might appear to be relevant to our experimental MP data.  \JA{However, our current TEMC experiments are stationary in the lab frame and, thus, are only spun by Earth's rotation, similar to \citet{brown2006effect} and unlike \citet{king2015magnetostrophic}. They therefore exist in the weakly rotating MC wave limit.  As shown in Appendix \ref{MCwave}, the weakly rotating MC wave dispersion relation can be reduced to} 
\be 
    \omega_{MC} = \pm \left( \frac{\bm{B} \boldsymbol \cdot \bm k_0}{\sqrt{\rho \mu}} \left( 1 \pm \epsilon \right) \right) \approx \omega_A, \quad \text{where} \quad 
    \epsilon = \frac{\sqrt{\rho \mu}}{k_0} \frac{\bm \Omega \boldsymbol \cdot \bm k_0}{\bm{B_0} \boldsymbol \cdot \bm k_0} \ll 1, 
\ee 
$\bm k_0$ is the wavevector, $\bm \Omega$ is the angular velocity vector, $\bm B$ is the applied magnetic flux density, $\mu$ is the magnetic permeability and $\omega_A = \pm(\boldsymbol{B \cdot k}_0)/\sqrt{\rho \mu}$ is the Alfv\'en wave dispersion relation. 
With $\epsilon \simeq 4 \times 10^{-6}$ s$^{-1}$ in our experiments, the MC wave period is well approximated by an Alfv\'en wave timescale $\tau_A = H/(B/\sqrt{\rho \mu})$ that is typically less than $1$ s in our system. For instance, $\tau_A \simeq 0.74$ s for the \blue{$Conducting\ MC^-$} case, a value three orders of magnitude shorter than the observed MP periods. Thus, we conclude that the MP mode is best explained via TEMC dynamics.

Turbulent magnetoconvection is relevant for understanding many geophysical and astrophysical phenomena \citep[e.g.,][]{proctor1982magnetoconvection,king2015magnetostrophic,vogt2021oscillatory}.  If $\widetilde{S} \nsim 0$ under planetary interior conditions \citep[e.g.,][]{chen2019enhancement}, then thermoelectric currents can exist in the vicinity of the core-mantle boundary (CMB) and inner core boundary of Earth's liquid metal outer core and those of other planets. Assuming $Se$ is not trivially small across these planetary interfaces, then TEMC-like dynamics could influence planetary core processes \JA{and could prove important for our understanding of planetary magnetic field observations \citep[e.g.,][]{merrill1977anomalies, stevenson1987mercury, schneider1988inclination, giampieri2002mercury, 
meduri2020numerical}.} 
Previous models of planetary core thermoelectricity have focused predominantly on magnetic fields produced as a byproduct of CMB thermoelectric current loops \citep{stevenson1987mercury,giampieri2002mercury}. In contrast, our experimental results suggest that TE processes can generate `slow modes,’ which could change a body's observed magnetic field by altering the local CMB magnetohydrodynamics. Further, thermoelectric effects provide a $\boldsymbol{B}$-dependent symmetry-breaker that does not exist in current models of planetary core magnetohydrodynamics.  

\begin{figure}
  \centering
    \includegraphics[width=0.7\textwidth]{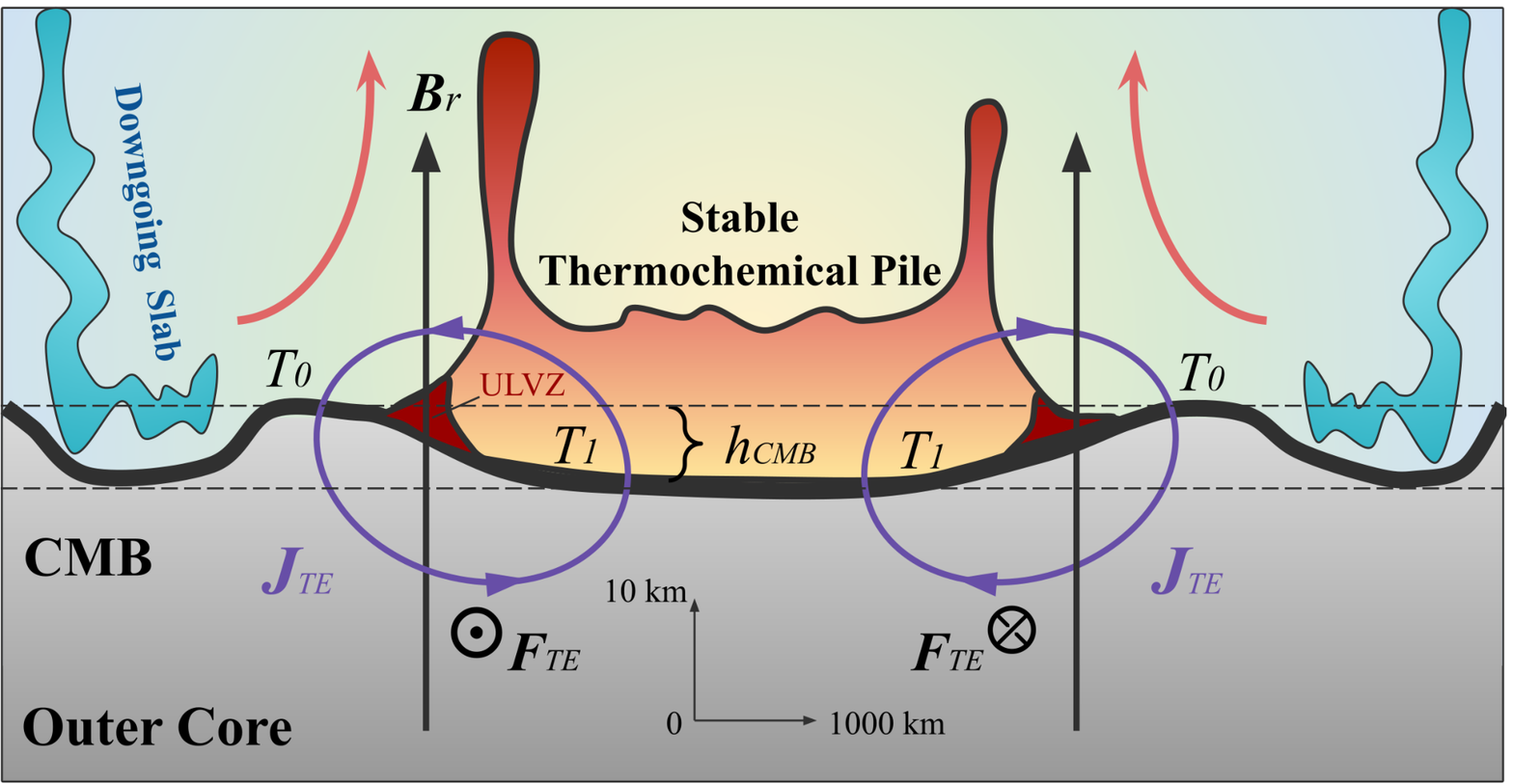}
    \caption{Schematic adapted from \cite{garnero2016continent} and \cite{deschamps2018constraints} showing thermoelectric currents $\boldsymbol J_{T\!E}$  and forces $\boldsymbol F_{T\!E}$ in the vicinty of the core-mantle boundary (CMB). The vertical scale is considerably exaggerated. The black arrows are radial magnetic field, $B_r \sim 19$ gauss. The temperature contrast between the thermochemical pile or the ULVZs and the surrounding mantle is denoted by $T_1-T_0=\Delta T_p \sim 10^2$ K. The dynamic depression of the CMB, $h_{C\!M\!B} \sim 5$ km, can generate smaller adiabatic temperature differences of order $5\ \mathrm K$.}
    \label{fig:cmb}
\end{figure}

Two dominant structures are known to exist at the base of Earth's mantle\JA{:} thermochemical piles \citep{trampert2004probabilistic,mosca2012seismic,garnero2016continent, deschamps2018constraints} and ultra low velocity zones (ULVZs) \citep{garnero1998ultralow}, both shown schematically in figure \ref{fig:cmb}. \blue{Seismic tomographic 
\JA{inversions reveal} that the continental-sized thermochemical piles have characteristic length scales of approximately $5000$ km along the CMB \citep{cottaar2016morphology,garnero2016continent}. The ULVZs are patches a few tens of kilometers thick just above the CMB where the seismic shear wave speed is about $30\%$ lower than the surrounding material \citep{garnero1998ultralow}. The lateral length scale of ULVZs \blue{ranges} from \JA{ $10$ km \citep{garnero2016continent} to over $1000$ km based on recent studies of so-called mega-ULVZs \citep{thorne2020new,thorne2021most}. Lateral thermal gradients can exist along the CMB between the piles or ULVZs and surrounding regions. By estimating the excess temperature of mantle plumes \citep{bunge2005low} or by taking the temperature difference between the ULVZ melt \citep[e.g.,][]{liu2016origins,li2017compositionally} and cold slabs \citep{tan2002slabs} near the CMB, we argue that the lateral temperature difference on the CMB to be $\lesssim 3 \times 10^2\ \mathrm K$.} Therefore, the lateral thermal gradient across the edge of ULVZs is possibly on the order of magnitude of $1\ \mathrm{K/km}$.}

\JA{On the fluid core side of the CMB, \cite{mound2019} have argued that the outer core fluid situated just below the thermochemical piles will tend to form regional stably-stratified lenses.} If such lenses exist, thermal gradients will also exist in the outer core across the boundaries between the stable lenses, where the heat flux is subadiabatic, and the surrounding convective regions. ULVZs may have high electrical conductivity due to iron enrichment and silicate melt \citep{holmstrom2018electronic}, such that the electrical conductivity is $\sigma \simeq 3.6 \times 10^4\ \mathrm{S/m}$ at CMB-like condition ($136\ \mathrm{GPa}\ \mathrm{and}\ 4000\ \mathrm K$). Therefore, these structures may prove well-suited to host TE current loops.

\begin{table}
    \centering
    \begin{tabular}{lll}
\textbf{Parameters} & \textbf{Description} & \textbf{Estimates} \\[5pt]
$\Delta T_p$        & Lateral temperature difference between the thermo-             & $\sim 300\ \mathrm{K}$         \\[3pt]
                    & chemical pile and its surrounding mantle in the CMB            &                                  \\[3pt]    
\JA{$U_{C}$ }           & Outer core flow speed near CMB                  & $\sim 0.1\ \mathrm{mm/s}$      \\[3pt]
$B_r$               & Radial geomagnetic field at the CMB                                & $\sim 1\ \mathrm{mT}$           \\[3pt]
$L_p$               & Characteristic length of a thermochemical pile                 & $\sim 5000 \ \mathrm{km}$     \\[3pt]
$L_{ulvz}$            & Characteristic length of the ULVZs                             & $\sim 500 \ \mathrm{km}$     \\[3pt]
    \end{tabular}
    \caption{\JA{Parameters used to estimate Seebeck numbers (\ref{equ:defSe3}) across Earth's core-mantle boundary (CMB).}}
    \label{tableCMB}
\end{table}
\begin{figure}
  \centering
    \includegraphics[width=0.5\textwidth]{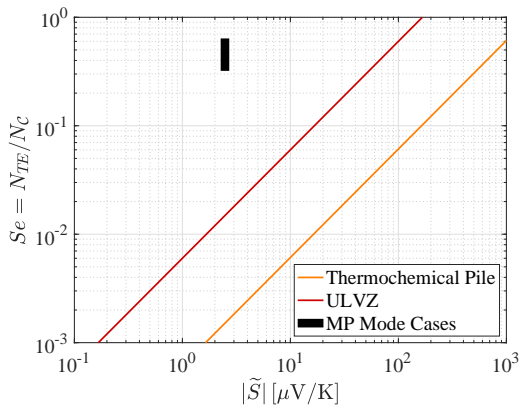}
    \caption{The Seebeck number $Se$ as a function of the net Seebeck coefficient $\widetilde S$. The red (orange) line represents estimated values for ULVZs (thermochemical piles) via eqs. (\ref{equ:defSe3}). The thick black line marks the range of cases that have magnetoprecessional modes in this study. Here, $\widetilde S$ is defined with the mean temperature of the fluid, $\overline T$, so that $\widetilde S = X_0 \overline T$. Our experiments show that TEMC dynamics can emerge at $Se$ below unity.}
    \label{fig:SvsSe}
\end{figure}

\JA{
In Earth's core, we take the radial magnetic field strength $B_r$ to be of order $1\, \mathrm{mT}$ on the CMB and estimate the flow speed to be $U_C \sim 0.1$ mm/s, based on inversions of geomagnetic field data \citep{holme2015large, finlay2016recent}. Note that the outer core flow velocity might be lower if there are convectively stable layers \citep{buffett2010stratification} or stable fluid lenses \citep{mound2019} situated below the CMB. Seebeck numbers across Earth's core-mantle boundary may then be estimated as
\begin{subequations}
    \begin{eqnarray}
    Se &=& \frac{|\widetilde{S}| \Delta T_p}{U_{C} B_r L_p} \ \text{for the thermochemical piles}, \\
    Se &=& \frac{|\widetilde{S}| \Delta T_p}{U_{C} B_r L_{ulvz}} \  \text{for the ULVZs},
    \end{eqnarray}
    \label{equ:defSe3}
\end{subequations}
where $\Delta T_p \sim 300\, \mathrm K$ is taken to be the lateral temperature difference between the thermochemical piles and the surrounding mantle. }

\JA{Figure \ref{fig:SvsSe} shows estimated CMB Seebeck numbers as a function of the net Seebeck coefficient $\widetilde S$. The orange line represents the thermochemical pile with a characteristic length $L_p \sim 5\times 10^3\, \mathrm{km}$. The red line represents the ULVZ with a characteristic length $L_{ulvz} \sim 500\, \mathrm{km}$. The cases from this study that have MP modes are denoted by \blue{a} thick black line. The Seebeck number is defined as $Se= |\widetilde S| \Delta T/(U_{f\!f}BH)$, with $\widetilde S$ defined with the mean temperature of the fluid, $\overline T$, so that $\widetilde S = X_0 \overline T$. Our experimental results show TEMC dynamics emerge at $Se$ below $1$. This plot suggests then that net Seebeck coefficients at the CMB must exceed values of order 10 $\mu$V/K in order for TEMC dynamics to affect Earth's local CMB dynamics and possibly alter the observed magnetic field. } 

\blue{We do not know at present if $\widetilde S$ can actually attain the values neccessary to drive significant thermoelectric currents across planetary core interfaces. It should be noted that many thermoelectric materials, especially semiconductors, are known to have large Seebeck coefficients that can exceed $100\, \mathrm{\mu V/K}$. Silicon, for instance, has a Seebeck coefficient of $\sim 800\, \mathrm{\mu V/K}$ at $500\, \mathrm{K}$ \citep{fulkerson1968thermal}. Moreover, recent studies in thermoelectric materials show that Seebeck coefficients can increase with increasing pressures and temperatures \citep{chen2019enhancement, morozova2019strategies, yoshino2020measurement}. How the TE coefficients of deep Earth materials extrapolate to core-mantle boundary conditions has, however, yet to be determined.}

\section*{Acknowledgements}
\JA{This manuscript was greatly improved thanks to feedback given by three anonymous referees. We further thank Ashna Aggarwal, Xiyuan Bao, Paul Davis, Carolina Lithgow-Bertelloni, Lars Stixrude, and Tobias Vogt for fruitful discussions, and Jewel Abbate and Taylor Lonner for help with the experiments. The authors also} gratefully acknowledge the support of the NSF Geophysics Program (EAR awards 1620649 and 1853196).
 
\section*{Declaration of Interests}
The authors report no conflict of interest.


%% file: sections/Appendix.tex
\label{AppA}

The governing equations (\ref{dimGE}) can be nondimensionalized using $H$ as length scale and $U_{f\!f}$ as the velocity scale, such that the free-fall time $\tau_{f\!f} = H / U_{f\! f}$ is the time scale. Moreover, the external magnetic flux density $B$ is the magnetic field scale, the bulk current density is scaled by $\sigma U_{f\!f} B$, the electric potential scale is $U_{f\!f} B H$, and $\Delta T$ is the temperature scale. 

The dimensionless governing equations of \JA{Oberbeck-Boussinesq} thermoelectric magnetoconvection (TEMC) are
\begin{subequations}
\label{nondimGE}
\begin{equation}
	 \boldsymbol \nabla  \boldsymbol \cdot  \boldsymbol { J } = 0,
	  \label{app:equ1}
\end{equation}	 
\begin{equation} 
	 \boldsymbol{J} = - \boldsymbol \nabla \Phi + \boldsymbol{u}\times \bm{\hat e_b} - Se \boldsymbol \nabla T, 
	  \label{app:equ2}
\end{equation}	
\begin{equation}
	 \boldsymbol \nabla  \boldsymbol \cdot  \boldsymbol { u } = 0,
	 \label{app:equ3}
\end{equation}	 
\begin{equation}
	 \frac{\partial  \boldsymbol{u}}{\partial {t}}+ \boldsymbol{u}  \boldsymbol \cdot {\boldsymbol \nabla }  \boldsymbol{u} = -{\boldsymbol \nabla } {p}+\sqrt{\frac{Ch^2 P r}{R a}} \left( \boldsymbol{J} \times \bm{\hat e_b}  \right) +\sqrt{\frac{P r}{R a}} {\boldsymbol \nabla }^{2}  \boldsymbol{u}+T \bm{\hat e_z}, 
	  \label{app:equ4}
\end{equation}	 
\begin{equation}
	 \frac{\partial T}{\partial {t}}+ \boldsymbol{u}  \boldsymbol \cdot {\boldsymbol \nabla } {T} = \sqrt{\frac{1}{R a P r}} {\boldsymbol \nabla }^{2} {T},
	  \label{app:equ5}
\end{equation}	 
\end{subequations}
where here $\boldsymbol{u}$ is the dimensionless velocity of the fluid, $p$ is the dimensionless non-hydrostatic pressure, $\boldsymbol{J}$ is the dimensionless electric current density, $\boldsymbol{B} = \bm{\hat e_b}$ is the dimensionless flux density of the external magnetic field, and $T$ is the nondimensional temperature. The nondimensionalized vertical magnetic field is constant and uniform, $\bm{\hat e_b} = \pm \bm{\hat e_z}$. 

The $(Ra/Pr)^{-1/2}$ grouping in (\ref{app:equ4}) is the reciprocal of the Reynolds number $Re$. The $(Ra Pr)^{-1/2}$ group in (\ref{app:equ5}) is the reciprocal of the P\'eclet number. Further, using (\ref{app:equ2}), the Lorentz term in (\ref{app:equ4}) expands to:
\begin{eqnarray}
    \sqrt{\frac{Ch^2 P r}{R a}} \left( \boldsymbol{J} \times \bm{\hat e_b}  \right)
    & = & \large( N_{\mathcal{C}}  \large[ - \boldsymbol \nabla \Phi + \boldsymbol{u}\times \bm{\hat e_b} \large] - \Nte \boldsymbol \nabla T\large) \times \bm{\hat e_b} \nonumber \\
    & = & N_{\mathcal{C}} \large[(\bm{\hat e_b} \times \boldsymbol \nabla \Phi )  - \boldsymbol{u}_\perp \large]  + \Nte (\bm{\hat e_b} \times \boldsymbol \nabla T) , 
    \label{nondimFL}
\end{eqnarray}
where $\boldsymbol{u}_\perp$ is the velocity perpendicular to the vertical direction $\boldsymbol{e}_z$. The first term on the right-hand side is due to irrotational electric fields in the fluid, which are likely small in our experiments. The second term is the quasi-static Lorentz drag, and the third term is due to thermoelectric currents in the fluid. The nondimensional groups in (\ref{nondimFL}) are the convective interaction parameter, $N_{\mathcal{C}} = Ch \sqrt{Pr/Ra}$, and the thermoelectric interaction parameter, $\Nte = Se \, N_{\mathcal{C}} $. 

%% file: sections/Appendix2.tex
The dispersion relation for the Magnetic-Coriolis wave is \citep{finlay2008course}
\begin{equation}
    \omega_{MC}=\pm \frac{\bm \Omega \boldsymbol \cdot \bm k_0}{k_0} \pm \left(\frac{\left(\bm \Omega \boldsymbol \cdot \bm k_0\right)^{2}}{k_0^{2}}+\frac{\left(\bm{B} \boldsymbol \cdot \bm k_0\right)^2}{\rho \mu}\right)^{1/2},
\end{equation}
where $\bm k_0$ is the wavenumber, $\bm \Omega$ is the angular rotation vector, and $\bm B$ is the magnetic flux density. In the weakly rotating limit, the Alfv\'en wave frequency, \JA{$\omega_A$,} is much larger than the inertial wave frequency: $| \boldsymbol{B\cdot k}/ \sqrt{\rho \mu}| \gg | 2 \boldsymbol{\Omega \cdot k}_0 /k_0 |$. The dispersion relation can then be rewritten as 
\begin{equation}
     \omega_{MC}=\pm \frac{\bm \Omega \boldsymbol \cdot \bm k_0}{k_0} \pm \frac{\bm {B} \boldsymbol \cdot \bm k_0}{\sqrt{\rho \mu}}\left(1+\frac{(\bm \Omega \boldsymbol \cdot \bm k_0)^{2} \rho \mu}{\left(\bm {B} \boldsymbol \cdot \bm k_0 \right)^{2} k_0^{2}}\right)^{1/2}.
    \label{equ:mcwave2}
\end{equation}
The last term of eq. (\ref{equ:mcwave2}) is small, allowing us to carry out a Taylor expansion, 
\begin{equation}
    \omega_{MC}  \JA{ = } \pm \frac{\bm \Omega \boldsymbol \cdot \bm k_0}{k_0} \pm \frac{\bm {B} \boldsymbol \cdot \bm k_0}{\sqrt{\rho \mu}}\left(1+\frac{(\bm \Omega \boldsymbol \cdot \bm k_0)^{2} \rho \mu}{2\left(\bm {B} \boldsymbol \cdot \bm k_0 \right)^{2} k_0^{2}}\right).
    \label{eq:MCB3}
\end{equation}
\blue{A small quantity $\epsilon$ can be defined as:
\begin{equation}
    \epsilon = \frac{\sqrt{\rho \mu}}{k_0} \frac{\bm \Omega \boldsymbol \cdot \bm k_0}{\bm{B} \boldsymbol \cdot \bm k_0}\ll 1.
\end{equation}
Eq. (\ref{eq:MCB3}) can then be recast with $\epsilon$,  
\begin{equation}
    \JA{\omega_{MC}} = \pm \frac{\bm{B} \boldsymbol \cdot \bm k_0}{\sqrt{\rho \mu}} \left( \epsilon \pm (1+\frac{1}{2} \epsilon^2) \right).
    \label{eq:MCB5}
\end{equation}
Two branches of solution emerge: the fast branch, $\omega_{MC}^f$, is acquired when the first two signs in eq. (\ref{eq:MCB5}) are the same. In contrast, the slow branch, $\omega_{MC}^s$, is acquired when the first two signs in eq. (\ref{eq:MCB5}) are the opposite. Therefore, the slow branch solution will have a smaller absolute values than the fast branch. The solution becomes
\begin{equation}
    \left.
    \begin{array}{l}
    \displaystyle
    \omega_{MC}^f = \pm \frac{\bm{B} \boldsymbol \cdot \bm k_0}{\sqrt{\rho \mu}} \left( \frac{1}{2} (1+\epsilon)^2 +\frac{1}{2} \right); \\[16pt]
    \displaystyle
    \omega_{MC}^s = \pm \frac{\bm{B} \boldsymbol \cdot \bm k_0}{\sqrt{\rho \mu}} \left( \frac{1}{2} (1-\epsilon)^2 +\frac{1}{2} \right).
    \end{array} \right\}
\end{equation}
We can further simplify this dispersion relation by applying another Taylor expansion, $(1\pm \epsilon)^2 \approx 1\pm 2\epsilon$}, such that:
\begin{equation}
    \left.
    \begin{array}{l}
    \displaystyle
    \omega_{MC}^f = \pm \frac{\bm{B} \boldsymbol \cdot \bm k_0}{\sqrt{\rho \mu}} \left( 1+        \epsilon \right) \approx \omega_A; \\[16pt]
    \displaystyle
    \omega_{MC}^s = \pm \frac{\bm{B} \boldsymbol \cdot \bm k_0}{\sqrt{\rho \mu}} \left( 1 -         \epsilon \right) \approx \omega_A.
    \end{array} \right\}
\end{equation}
Thus, both fast and slow Magnetic-Coriolis waves \JA{behave as} Alfv\'en waves in the weakly rotating limit.

%% file: sections/Appendix3.tex
\begin{figure}
    \centering
    \includegraphics[width=\linewidth]{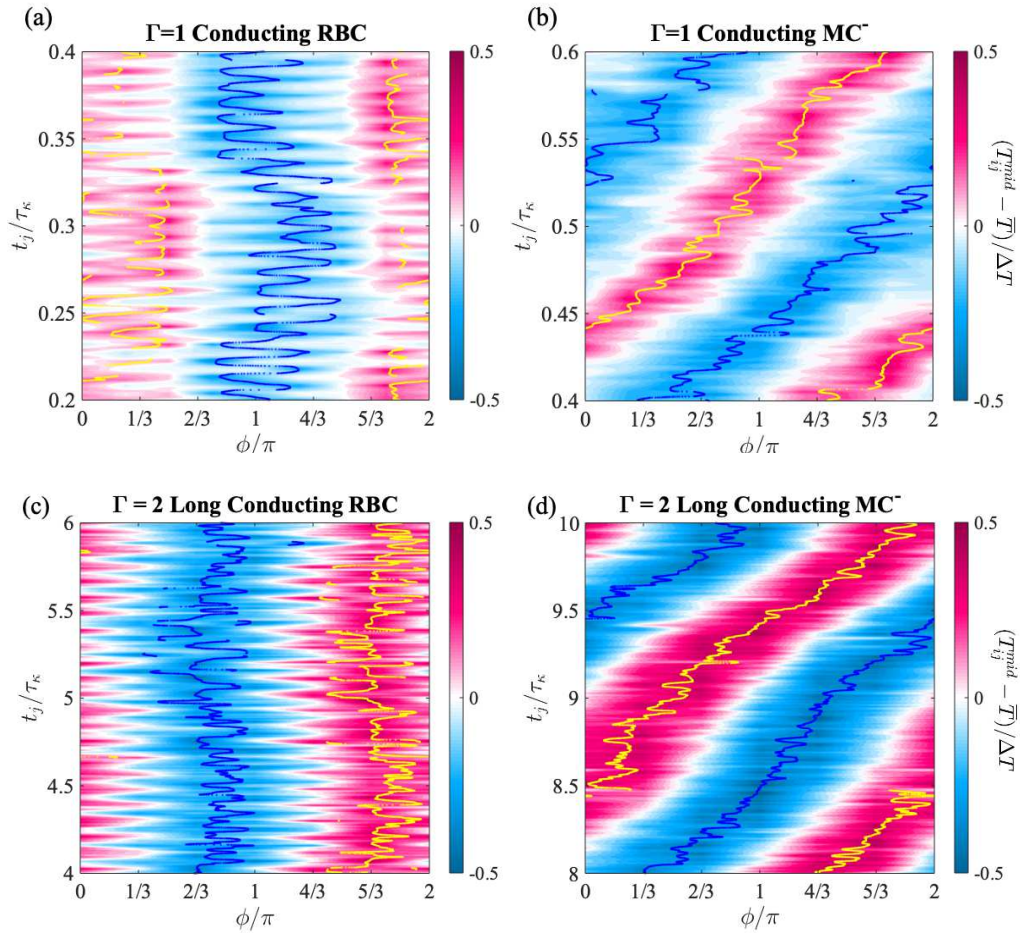}
    \caption{\SH{Hovm\"oller diagrams of the sidewall midplane temperature fluctuation. (a)~$Conducting\ RBC$ case in the $\Gamma = 1$ tank at $Ra = 2\times 10^7$, showing the zig-zag pattern characteristic for sloshing. (b) Corresponding $Conducting\ MC^-$ case at ($Ra = 2.8\times 10^7,\, Ch = 4.1\times 10^4$; \JA{$N_\mathcal{C} = 1.27,\, \Nte = 0.16 $}) showing a drifting magnetoprecession mode. (c) $Long\ Conducting\ RBC$ case in $\Gamma = 2$ tank at $Ra = 1.79 \times 10^6$ (same as in the main manuscript), showing the accordion pattern characteristic for a jump rope vortex (JRV). (d) Corresponding  $Long\ Conducting\ MC^-$ at ($Ra = 1.82 \times 10^6, \, Ch = 2.6 \times 10^3$ ; \JA{$N_\mathcal{C} = 0.31,\, \Nte = 0.16 $}), showing a drifting magnetoprecession mode similar to (b). \JA{The} yellow (blue) lines indicate the position of the maximum (minimum) temperature obtained with the TEE method. The time windows are selected to show approximately one full precession, corresponding, respectively, to $0.2 \tau_\kappa$ for $\Gamma = 1$ and to $2 \tau_\kappa$ for $\Gamma = 2$.}} 
    \label{fig:contours}
\end{figure} 

\blue{To broaden our understanding of TEMC dynamics, here we compare and contrast a set of $\Gamma = 1$ experiments with comparable $\Gamma = 2$ cases. Using the same setup as described in the main text, the $H \simeq 10$ cm sidewall ($\Gamma = 2$) is exchanged with an $H \simeq 20$ cm sidewall to create an experimental device with a $\Gamma = 1$ tank geometry. } 

\blue{Figure \ref{fig:contours} shows Hovm\"oller plots of the sidewall temperature field in four separate experimental cases, one RBC case and one $Conducting \ MC^-$ case in each geometry.  Figure \ref{fig:contours}(b) shows that the thermal field precesses in the $\Gamma = 1$ $Conducting \ MC^-$ case, similarly to the MP behavior found in $\Gamma =2$ experiments. The $\Gamma = 1$ magnetoprecession rate is $d \xi^{mid}_j / dt = 8.40 \times 10^{-3}$ rad/s. Our model predicts the angular frequency of magnetoprecession to be $\omega_{MP} (U_{MC}/U_{f\!f}) = 6.22 \times 10^{-3}$ rad/s, which is in good zeroth order agreement with the data. }

\JA{The biggest difference then between the $\Gamma = 1$ and 2 cases is found to be the internal oscillation mode of the LSC.}
\SH{We have previously demonstrated in convection cells with $\Gamma = 2$ that the dominant LSC mode is a jump rope vortex (JRV), whereas in $\Gamma = 1$ the dominant mode is the coupled sloshing-torsional mode \citep{vogt2018jump}, see especially the Supplementary Material. \citet{brown2009origin} and \citet{zhou2009oscillations} have shown that the sloshing and torsional mode are contained in the very same advected oscillation. Therefore, both modes  have the same frequency but a phase difference of $\pi$, and the torsional mode cannot exist without a sloshing mode.}

\blue{We have verified this behaviour in our current set-up. The Conducting RBC case in $\Gamma = 1$ (figure~\ref{fig:contours}(a)) shows a zigzag pattern, which is characteristic for sloshing. The Conducting RBC case in $\Gamma =2$ (figure~\ref{fig:contours}(c)) shows an accordion pattern, which is characteristic for the jump rope vortex (JRV). In \citet{vogt2018jump}, we have demonstrated that this is the most straightforward method to identify either mode. In the magnetoconvection cases (figure~\ref{fig:contours}(b,d)), both the sloshing and the JRV modes  are suppressed by the magnetoprecession mode, which manifests itself through a strong azimuthal drift of the temperature pattern. But there are faint indications in the MC cases that a much weakened sloshing mode still exists in $\Gamma = 1$ cases, and similarly a much weakened JRV mode still exists in $\Gamma = 2$ cases. The latter is further supported by the spectral data shown in figure \ref{fig:spg}(a).} 

%
%

\blue{We have analysed our data more quantitatively by applying the TEE method of \citet{zhou2009oscillations} to the sidewall midplane, top, and bottom temperature measurements, as also indicated by the yellow and blue lines in figure~\ref{fig:contours}. These TEE measurements confirm the existence of the torsional mode in our $\Gamma = 1$ tank by measuring the difference in the azimuthal angles of the best-fit extrema between top and bottom (not shown). 
We did not find clear evidence of torsional oscillations in the $\Gamma = 2$ cases, neither in the $Insulating$ and $Conducting\ RBC$ cases nor in the $Insulating\ MC^\pm$, in agreement with the previous $\Gamma = 2$ results of \citet{vogt2018jump}. }

%% file: sections/data.tex
\label{AppC}
\begin{landscape}
    \begin{table}
        \begin{center}
            \begin{tabular}{lccccccccccccc}
$ {Cases}$  & ${Bi}$ & ${B} (\mathrm{Gs})$ & $\overline{T} (^{\circ}\mathrm{C})$  & ${\Delta T} (\mathrm{K})$ & ${P}(\mathrm{W})$    & ${Ra}(\!\times10^{-6}\!)$     & ${Ch}(\!\times10^{-3}\!)$       & ${N_\mathcal{C}}$   & \blue{${\Nte}$}  & ${Rm}(\!\times10^{2}\!)$   & ${Nu}$    & ${f/f_{\kappa}}$ &${t/\tau_\kappa}$\\
\hline\\ [-1.5ex]
$Insulating\ MC^{-}$    & 0.24      & -120   & 42.86       & 6.79       & 393.26    & 1.61      & 2.42        & 0.31       & 0         & 1.31       & 5.79$\pm$0.09     &  -        & 13.6\\
$Insulating\ RBC$       & 0.24      & 0      & 42.86       & 6.83       & 393.26    & 1.61      & 0           & 0            & 0       & 1.31       & 5.76$\pm$0.09     &  -        & 15.2\\
$Insulating\ MC^{+}$    & 0.24      & 120    & 42.85       & 6.79       & 393.24    & 1.61      & 2.41        & 0.31       & 0         & 1.31       & 5.79$\pm$0.09     &  -        & 14.0\\[1ex]
\hline\\ [-1.5ex]
$Conducting\ MC^{+}$    & 0.14      & 120   & 42.47         & 7.02       & 396.07    & 1.83      & 2.57        & 0.31      & 0.16       & 1.40       & 5.83$\pm$0.07     & 0.71     & 15.3\\
$Conducting\ RBC$       & 0.15      & 0     & 42.45         & 6.91       & 396.12    & 1.80      & 0           & 0           & 0        & 1.39       & 5.93$\pm$0.09     & 11.02    & 17.0\\
$Conducting\ MC^{-}$    & 0.14      & -121  & 42.50         & 7.03       & 396.04    & 1.83      & 2.59        & 0.31      & 0.16       & 1.40       & 5.82$\pm$0.08     & 0.61     & 15.8\\[1ex]
\hline\\[-1.5ex]
$Long\ Conducting\ MC^{+}$  & 0.14  & 120   & 42.45         & 7.01       & 396.16    & 1.82      & 2.58        & 0.31      & 0.16       & 1.40       & 5.84$\pm$0.07     & 0.67     & 37.4\\
$Long\ Conducting\ RBC$     & 0.15  & 0     & 42.41         & 6.88       & 396.19    & 1.79      & 0           & 0           & 0        & 1.39       & 5.96$\pm$0.09     & 11.07    & 36.1\\
$Long\ Conducting\ MC^{-}$  & 0.14  & -120  & 42.44         & 6.99       & 396.16    & 1.82      & 2.59        & 0.31      & 0.16       & 1.40       & 5.86$\pm$0.08     & 0.64     & 40.1\\[1ex]
\hline\\ [-1.5ex]
\JA{$\Gamma = 1\ Conducting\ RBC$}    & 0.16    & 0     & 44.47     & 9.89      & 627.83    & 20.57     & 0        & 0          & 0       & 4.68      & 13.10$\pm$0.13   & 109.52   & 2.4 \\
\JA{$\Gamma = 1\ Conducting\ MC^{-}$} & 0.16   & -234  & 52.79     & 13.00     & 835.49    & 28.06     & 40.92    & 1.23     & 0.16    & 5.36    & 13.24$\pm$0.10        & 4.59     & 3.7\\[1ex]
\hline\\ [-1.5ex]
    \end{tabular}
    \caption{\blue{Data and parameters of nine long-period $MC^+$, $RBC$, and $MC^-$ experiments in a $\Gamma = 2$ cell with electrically conducting copper end-blocks and electrically insulating Teflon end-blocks, respectively; and two $\Gamma = 1\ Conducting\ RBC$ and $MC^-$ cases. In all the cases above, $Pm = 1.69 \times 10^{-6}$, and $Pr = 0.0267$. The precessional frequencies for both long/short groups differ by $\lesssim 5\%$. The Biot number $Bi$ shown in this table is the average Biot number of the top and bottom thermal blocks. $B$ denotes the magnetic flux density; $\overline{T}$ is the time-averaged temperature of the top and bottom plates; $\Delta T$ is the temperature difference across the tank height; $P$ is the net power input with heat loss correction; $Ra$ is the Rayleigh number; $Ch$ is the Chandrasekhar number; $\Nc$ is the convective interaction parameter; $\Nte$ is the thermoelectric interaction parameter, defined in eq. (\ref{equ:defNTE}); $Rm$ is the free-fall estimate of magnetic Reynolds number; $Nu$ is the Nusselt number; $f/f_{\kappa}$ is the peak frequency of the averaged spectrum over six sidewall thermistors, normalized by the thermal diffusion frequency; and $t/\tau_\kappa$ denotes the time duration of each case in thermal diffusion timescale $\tau_\kappa=H^2/\kappa$, after the time series data reached statistical equilibrium. Each case is equilibrated for at least $30$ min ($ > 2 \tau_\kappa$). The long conducting cases lasted for about $\approx 113$ thermal diffusion time, whereas the short conducting cases lasted for $\approx 48$ thermal diffusion time. The insulating cases lasted for $\approx 43$ thermal diffusion time. } }
            \label{app:table1}
        \end{center}
    \end{table}
    
    
    \begin{table}
        \begin{center}
            \begin{tabular}{cccccccccccccccc}
${B} (\mathrm{Gs})$  & $\overline T (^{\circ}\mathrm{C})$  & ${\Delta T } (\mathrm{K})$ & ${P}(\mathrm{W})$    & ${Ra}(\!\times\!10^{-6}\!)$   & ${Ra/Ra^{\infty}_{A}}$ &  ${Ra/Ra^{\infty}_{W}}$  & ${Ch}$       & ${N_\mathcal{C}}$   & \blue{$\Nte$}       & ${Re}(\!\times\!10^{-3}\!)$    & ${Rm}(\!\times10^{2}\!)$    & ${Nu}$ & ${ f/f_{\kappa}}$   & ${f_{jrv}/f_{\kappa}}$   & ${t/\tau_\kappa}$   \\[0.5ex]
            \hline\\[-1.5ex]
0        & 40.99 & 8.19  & 444.78     & 2.12 &  -       &    -    & 0.00                & 0.00      & 0.00       & 8.85     & 1.51   & 5.61$\pm$0.17 & 11.34    & 11.34    & 53.43\\
-12      & 40.59 & 7.91  & 424.72     & 2.04 & 7945.81  & 710.06  & 26.00               & 0.003     & 0.02       & 8.72     & 1.49   & 5.55$\pm$0.17 & 11.19    & 11.19    & 45.41\\
-30      & 42.20 & 7.77  & 416.10     & 2.02 & 1238.28  & 242.21  & 1.66$\times 10^2$   & 0.02      & 0.04       & 8.67     & 1.47   & 5.53$\pm$0.16 & 11.11    & 11.11    & 66.80\\
-50      & 42.22 & 7.77  & 416.09     & 2.02 & 455.96   & 133.53  & 4.50$\times 10^2$   & 0.05      & 0.07       & 8.67     & 1.47   & 5.53$\pm$0.15 & 10.98    & 10.98    & 53.44\\
-80      & 40.71 & 7.44  & 393.93     & 1.92 & 172.20   & 72.23   & 1.13$\times 10^3$   & 0.13      & 0.10       & 8.43     & 1.44   & 5.47$\pm$0.10 & 10.66    & 10.66    & 28.05\\
-101     & 40.67 & 7.52  & 393.95     & 1.94 & 110.05   & 54.92   & 1.79$\times 10^3$   & 0.21      & 0.13       & 8.49     & 1.45   & 5.41$\pm$0.07 & 0.24     & 9.56     & 46.75\\
-120     & 40.73 & 7.64  & 394.08     & 1.97 & 77.98    & 44.50   & 2.57$\times 10^3$   & 0.30      & 0.16       & 8.55     & 1.46   & 5.33$\pm$0.08 & 0.43     & 8.82     & 66.78\\
-137     & 40.69 & 7.81  & 394.33     & 2.02 & 61.23    & 38.50   & 3.34$\times 10^3$   & 0.39      & 0.18       & 8.66     & 1.48   & 5.22$\pm$0.07 & 0.56     & -        & 53.43\\
-151     & 40.85 & 7.88  & 393.83     & 2.04 & 51.31    & 34.52   & 4.02$\times 10^3$   & 0.46      & 0.20       & 8.73     & 1.49   & 5.17$\pm$0.07 & 0.60     & -        & 60.11\\
-170     & 40.91 & 7.97  & 393.80     & 2.06 & 40.94    & 30.01   & 5.10$\times 10^3$   & 0.58      & 0.22       & 8.73     & 1.49   & 5.10$\pm$0.13 & 0.49     & -        & 66.79\\
-201     & 40.98 & 7.43  & 364.19     & 1.92 & 27.26    & 22.53   & 7.15$\times 10^3$   & 0.85      & 0.26       & 8.44     & 1.44   & 5.07$\pm$0.20 & 0.31     & -        & 86.82\\
-241     & 41.04 & 7.50  & 364.14     & 1.94 & 19.17    & 17.99   & 1.03$\times 10^4$   & 1.21      & 0.31       & 8.50     & 1.45   & 5.02$\pm$0.10 & 1.56     & -        & 32.06\\
-270     & 41.18 & 7.82  & 364.44     & 2.02 & 15.87    & 16.13   & 1.29$\times 10^4$   & 1.49      & 0.35       & 8.68     & 1.48   & 4.82$\pm$0.08 & 1.59     & -        & 66.79\\
-301     & 41.30 & 8.06  & 364.02     & 2.09 & 13.18    & 14.43   & 1.60$\times 10^4$   & 1.82      & 0.39       & 8.81     & 1.50   & 4.67$\pm$0.09 & 1.18     & -        & 46.75\\
-447     & 39.96 & 8.24  & 296.14     & 2.12 & 6.09     & 8.69    & 3.53$\times 10^4$   & 3.99      & 0.58       & 8.87     & 1.52   & 3.71$\pm$0.05 & 1.33     & -        & 46.74\\
-598     & 38.79 & 8.17  & 245.93     & 2.09 & 3.38     & 5.82    & 6.26$\times 10^4$   & 7.15      & 0.78       & 8.76     & 1.51   & 3.11$\pm$0.04 & 0.90     & -        & 53.41\\
-693     & 38.39 & 7.72  & 196.48     & 1.97 & 2.38     & 4.50    & 8.41$\times 10^4$   & 9.89      & 0.90       & 8.51     & 1.47   & 2.63$\pm$0.03 & 0.53     & -        & 100.15\\
\hline\\ [-1.5ex]
            \end{tabular}
        \caption{Data and output parameters from the study of the precessional modes' evolution. In all the cases above, $\Gamma =2,\ Pm = 1.71 \times 10^{-6}$, and $Pr = 0.025$. \blue{In this table, $B$ is the magnetic flux density; $\overline T$ is the average temperature of the top and bottom plates; $\Delta T$ is the temperature drop across the tank height; $P$ is the net power input with heat loss correction; $Ra^{\infty}_{A}$ and $Ra^{\infty}_{W}$ are critical $Ra$ predictions from Chandrasekhar's asymptotic solution for an infinite plane \citep{chandrasekhar1961hydrodynamic} and Busse's asymptotic solution for a semi-infinite plane with a vertical wall \citep{busse2008asymptotic}; $Ch$ is the Chandrasekhar number; $\Nc$ is the convective interaction parameter; $\Nte$ is the thermoelectric interaction parameter, defined in eq. (\ref{equ:defNTE}); $Rm$ is the free-fall estimate of the magnetic Reynolds number; $Nu$ is the Nusselt number; $ f/f_{\kappa}$ is the peak frequency of the averaged spectrum over six sidewall thermistors, normalized by thermal diffusion frequency; $f_{jrv}/f_{\kappa}$ is the jump-rope vortex frequency, normalized by thermal diffusion frequency; and $t/\tau_\kappa$ is time normalized by $\tau_\kappa = H^2/\kappa$, after the time series data reached statistical equilibrium. Prior to the acquisition of the data, each case is equilibrated for at least $30$ min, which is over $2 \tau_\kappa$.}}
        \label{app:table2}
        \end{center}
    \end{table}

\end{landscape}